\documentclass[twocolumn]{aastex631}

\newcommand\aastex{AAS\TeX}
\newcommand\latex{La\TeX}
\newcommand{\kms}{km\,s$^{-1}$}
\newcommand{\Kkmspc}{K km\,s$^{-1}$ pc$^{2}$}

\newcommand{\jybmkms}{Jy~bm$^{-1}$\,$\times$\,km\,s$^{-1}$}
\newcommand{\SQ}{Stephan's Quintet}
\newcommand{\HI}{H\,{{\small \sc I}}}
\usepackage{hyperref}

\begin{document}

\title{Bird's-eye View of Molecular Gas across Stephan's Quintet Galaxy Group and Intra-group Medium}
\shorttitle{Cold Molecular Gas across Stephan's Quintet}
\shortauthors{Emonts et al.}

\correspondingauthor{Bjorn Emonts}
\email{bemonts@nrao.edu}

\author[0000-0003-2983-815X]{B.\,H.\,C. Emonts}
\affiliation{National Radio Astronomy Observatory, 520 Edgemont Road, Charlottesville, VA 22903, USA}

\author[0000-0002-7607-8766]{P.\,N. Appleton}
\affiliation{Caltech/IPAC, MC 314-6, 1200 E. California Blvd., Pasadena, CA 91125, USA.}

\author[0000-0002-9471-5423]{U. Lisenfeld}
\affiliation{Departamento de F\'{i}sica Te\'{o}rica y del Cosmos, Universidad de Granada, 18071 Granada, Spain}
\affiliation{Instituto Carlos I de F\'{i}sica Te\'{o}rica y Computacional, Facultad de Ciencias, 18071 Granada, Spain}

\author[0000-0002-2421-1350]{P. Guillard}
\affiliation{Sorbonne Universit\'{e}, CNRS, UMR 7095, Institut d'Astrophysique de Paris, 98bis bd Arago, 75014 Paris, France}
\affiliation{Institut Universitaire de France, Minist\`{e}re de l'Enseignement Sup\'{e}rieur et de la Recherche, 1 rue Descartes, 75231 Paris Cedex 05, France}

\author[0000-0002-1588-6700]{C.\,K. Xu}
\affiliation{Chinese Academy of Sciences South America Center for Astronomy, National Astronomical Observatories, CAS, Beijing 100101,
People's Republic of China}
\affiliation{National Astronomical Observatories, Chinese Academy of Sciences (NAOC), 20A Datun Road, Chaoyang District, Beijing 100101, People's Republic of China
}

\author[0000-0001-8362-4094]{W.\,T. Reach}
\affiliation{Space Science Institute, 4765 Walnut St, Suite B, Boulder, CO 80301.}

\author[0000-0003-0057-8892]{L. Barcos-Mu\~{n}oz}
\affiliation{National Radio Astronomy Observatory, 520 Edgemont Road, Charlottesville, VA 22903, USA}

\author[0000-0002-0690-8824]{A. Labiano}
\affiliation{Telespazio UK S.L. for the European Space Agency (ESA), ESAC, Spain.}

\author[0000-0002-3471-981X]{P.\,M. Ogle}
\affiliation{Space Telescope Science Institute, 3700 San Martin Dr., Baltimore, MD 21218}

\author[0000-0002-5671-6900]{E. O'Sullivan}
\affiliation{Center for Astrophysics $|$ Harvard $\&$ Smithsonian, 60 Garden Street, Cambridge, MA, 02138, USA}

\author[0000-0001-5042-3421]{A. Togi}
\affiliation{Texas State University, 601 University Dr, San Marcos, TX 78666, USA}

\author[0000-0001-6217-8101]{S.\,C. Gallagher}
\affiliation{Institute for Earth and Space Exploration, Western University, 1151 Richmond St., London, ON N6A 3K7, Canada}

\author[0009-0001-2178-4022]{P. Aromal}
\affiliation{Institute for Earth and Space Exploration, Western University, 1151 Richmond St., London, ON N6A 3K7, Canada}

\author[0000-0003-3343-6284]{P.-\,A. Duc}
\affiliation{Universit\'{e} de Strasbourg, CNRS, Observatoire astronomique de Strasbourg (ObAS), UMR 7550, 67000 Strasbourg, France}

\author[0000-0002-4261-2326]{K. Alatalo}
\affiliation{Space Telescope Science Institute, 3700 San Martin Dr., Baltimore, MD 21218}
\affiliation{Johns Hopkins University, Department of Physics and Astronomy, Baltimore, MD 21218, USA}

\author[0000-0003-1097-6042]{F. Boulanger}
\affiliation{UPMC Universite Paris 06, \'{E}cole Normale Sup\'{e}rieure, 75005 Paris, France}

\author[0000-0003-0699-6083]{T. D\'{i}az-Santos}
\affiliation{Institute of Astrophysics, Foundation for Research and Technology-Hellas (FORTH), Heraklion, 70013, Greece.}
\affiliation{School of Sciences, European University Cyprus, Diogenes street, Engomi, 1516 Nicosia, Cyprus.}

\author[0000-0003-3367-3415]{G. Helou}
\affiliation{IPAC, California Institute of Technology, 1200 E. California Blvd, Pasadena, CA 91125, USA}




\begin{abstract}

We present the large-scale distribution and kinematics of cold molecular gas across the compact galaxy group \SQ, based on CO(2-1) observations performed with the Atacama Compact Array (ACA) and CO(1-0) data from the Combined Array for Research in Millimeter-wave Astronomy (CARMA). We find coherent structures of molecular gas associated with the galaxies and intra-group medium, which follow the distribution of warm H$_{2}$ previously seen with the James Webb Space Telescope ({\it JWST}). CO is associated with a ridge of shocked gas that crosses the galaxy group, and with a spiral arm of the intruding galaxy NGC\,7318b, which interacts with the intra-group medium along the ridge. Although the ridge contains widespread shocks, turbulent gas, and warm H$_{2}$, the CO lines are narrower than elsewhere in \SQ\ (FWHM\,$\sim$\,25\,$-$\,65 \kms), indicative of settled cold gas. At a distinctly different velocity, CO is found in the active galaxy NGC\,7319 and Northern star-forming region SQ-A. A bridge of turbulent molecular gas connects NGC\,7319 with the ridge, covering a gap of $\sim$700 \kms\ between these structures. The gas excitation ranges from $L_{\rm CO(2-1)}^{\prime}$/$L_{\rm CO(1-0)}^{\prime}$\,$\sim$\,0.3 in the bridge and SQ-A, to $\sim$0.5 along the ridge, to  near unity in the center of NGC\,7319. We also detect either a molecular outflow or turbulent molecular gas associated with the radio source in NGC\,7319. These ACA data are part of a program with the Atacama Large Millimeter/submillimeter Array (ALMA) and {\it JWST} to study molecular gas physics from the largest to the smallest scales across the intra-group medium of \SQ.
\end{abstract}



\section{Introduction} \label{sec:intro}

Stephan's Quintet is a compact group of galaxies also known as Hickson Compact Group 92 (HCG92; \citealt{hickson82}). It consists of four galaxies that are at a similar redshift of z\,$\sim$\,0.02 (NGC\,7317, 7318a, 7318b, and 7319), plus a foreground galaxy at $z$\,=\,0.0026 (NGC\,7320). These galaxies are often portrayed in iconic imagery of this system, such as the Early Release imaging with the James Webb Space Telescope ({\it JWST}; \citealt{pontoppidan22}).\footnote{\url{https://webbtelescope.org/contents/news-releases/2022/news-2022-034}} The main group contains an additional fainter galaxy NGC\,7320c at a similar redshift, which lies further away to the East of the main group. This galaxy appears at the tip of a prominent tidal tail of neutral hydrogen gas and stars, which originates at NGC\,7319 \citep{williams02,duc18}. 

In terms of the group dynamics, galaxy NGC\,7318b is likely `invading' \SQ\ on a first-time passage into the group \citep[e.g.,][]{moles98}, as it is approaching from behind at a blueshifted velocity of $\sim$1000 km$^{-1}$ compared with the other group members, which have a luminosity-weighted barycentric group velocity of 6600 \kms\ \citep{guillard22}. This barred spiral galaxy (SBc) with loosely wound spiral arms forms an apparent pair with the elliptical galaxy NGC\,7318a. The fainter early-type galaxy NGC\,7320c, which lies on the outskirts of the group, passed through the system at an earlier stage as part of an interaction with NGC\,7319 \citep[][]{moles98,renaud10,hwang12}. NGC\,7319 is a SBc galaxy with distorted morphology, which also contains a Seyfert-2 Active Galactic Nucleus (AGN) with a radio source \citep{sulentic01, aoki99, xanthopoulos04, pereira22}. Tidal and collisional stripping of gas removed much of the neutral hydrogen (\HI) gas from the interacting galaxies, given that the bulk of \HI\ gas is found in the intra-group medium and tidal debris of \SQ\ \citep{allen80,shostak84,williams02,xu22}. NGC 7317, an elliptical galaxy (E4) that lies in the South-West of the group, does not show any obvious signs of interaction with the other galaxies, except for a possible link to NGC 7319 by a diffuse optical and X-ray halo \citep{moles98,trinchieri05,duc18}. The optical halo resembles the intracluster light seen in clusters and fossil groups, and could hint to a group formation several Gyr ago \citep{duc18}.

A spectacular structure in \SQ\ is a shock-front that stretches midway between the interacting galaxies. This shock-front is detected as a `ridge' of emission across a wide range of the electromagnetic spectrum, from the X-rays \citep{trinchieri03,osullivan09} to the radio continuum \citep{allen72,xu03}, It contains regions of star formation \citep{xu99,gallagher01,konstantopoulos14}, but also regions where shocks inhibit star formation \citep{cluver10,konstantopoulos14}. This ridge is strongly multi-phase in nature, as it is detected in ultra-violet and optical lines of ionized gas that show a high velocity dispersion \citep[e.g.,][]{iglesias12,rodriguez14,guillard22}, in the mid-infrared (IR) H$_2$ lines of warm molecular gas detected with {\it Spitzer} and {\it JWST}~ \citep{appleton06, appleton17, appleton23, cluver10}, and in [CII] and H$_2$O emission observed with the {\it Herschel Space Observatory} \citep{appleton13}. It has also been observed in CO emission of cold molecular gas with the 30m single-dish telescope of the Institut de Radioastronomie Millim\'{e}trique \citep[IRAM;][]{guillard12,yttergren21} and radio interferometer of the Berkeley Illinois Maryland Association \citep[BIMA;][]{gao00}. Shock models, combined with {\it JWST} early-release imaging and observations of certain regions with ALMA, revealed that the molecular gas in the ridge underwent dissipation of mechanical energy along an elongated shock-front, which was created by the interaction among galaxies and the intra-group medium \citep{guillard09, appleton23}. 

Perpendicular to the ridge is a structure called the ‘bridge’, which extends towards NGC\,7319 in the West. It was detected in warm H$_{2}$ \citep{cluver10, appleton17}, CO \citep{guillard12}, [CII] \citep{appleton13} and H$\beta$ \citep{guillard22}. The gas in the bridge likely experiences strong turbulence, as indicated by broad line profiles in single-dish CO spectra \citep{guillard12} and high levels of heating of the warm H$_{2}$ \citep{appleton17}.

Along the extension of the ridge at the Northern end, but kinematically distinct from the ridge, lies a prominent extragalactic star-forming region \citep{xu99}, which has been studied in visible light using {\it HST} \citep{gallagher01, fedotov11}. This star-forming region, like the main ridge, has a counterpart in radio continuum \citep{xanthopoulos04} and molecular gas \citep{smith01,lisenfeld02,gao00,guillard12}. It was previously referred to as `source A' \citep{xu99}, `region A' \citep{smith01}, or `SQ-A' \citep[][hereafter in this paper]{lisenfeld02}.

In this paper, we present new observations of CO(2-1) performed with the Atacama Compact Array (ACA), which trace the large-scale distribution and kinematics of cold molecular gas in \SQ. We combine this with archival CO(1-0) data obtained with the Combined Array for Research in Millimeter-wave Astronomy (CARMA) to estimate the CO line ratios and molecular gas masses. Given that cold molecular gas is the raw material for star formation, a coherent view of the large-scale molecular gas properties is critical for our understanding of the evolution of \SQ. Our data distinguish the regions of the shock-front, the inner galaxies (NGC\,7318a/b and NGC\,7319), and SQ-A with much better detail than previous single-dish observations \citep{smith01,lisenfeld02,lisenfeld04,guillard12,yttergren21}. As such, these ACA data provide, for the first time with uniform sensitivity, a global overview of the molecular gas across the central $\sim$70$\times$70 kpc$^{2}$ of \SQ. Our goals for this paper are to study the global distribution, kinematics, mass, and excitation conditions of the cold molecular gas in the galaxies and the intra-group medium. These observations are part of an ongoing observational program with ALMA (program 2023.1.00177.S; PI Appleton) and {\it JWST} (Cycle-2 program GO-3445; PI Appleton). Following a pilot study by \citet{appleton23}, the ultimate goal of our larger ALMA/{\it JWST} program, once future high-resolution ALMA data will be added, is to investigate the physics of both cold and warm molecular gas from the largest to the smallest scales across \SQ, with a special focus on understanding the dissipation of energy across the shocked intra-group medium.

We assume a distance to \SQ\ of 94 Mpc, for which 1 arcsec corresponds to a spatial scale of 0.46 kpc. The redshift is $z$\,=\,0.0215 \citep{hickson82}.

\section{Data}

\subsection{Atacama Compact Array (ACA)}
\label{sec:acadata}
The main thrust of this paper is based on millimeter observations of Stephan's Quintet with the ACA, which is an array of 7m dishes that is part of ALMA. The observations were performed during 8$-$27 October 2023. The total on-source time was 10 hours, divided across 15 pointings, which were Nyquist sampled in a hexagonal mosaic pattern. The effective integration time for each area of the mosaic corresponds to that obtained in a single pointing of 1.7h. Observations were performed with two sets of two contiguous spectral windows, with each set covering $\sim$3.6 GHz with 1 MHz channels. One set of spectral windows was centred at 225.6\,GHz and captured the emission of the redshifted CO(2-1) line ($\nu_{\rm rest}$\,=\,230.54\,GHz). The other set of spectral windows was centred on 240.0\,GHz and captured continuum emission. The Largest Angular Scale on which these ACA observations can detect extended emission in a single channel is $\sim$30 arcsec ($\sim$14 kpc). The limited {\sl (u,v)}-coverage of the ACA resulted in low-level artifacts near strong emission-line features, which could mean that our analysis might miss very faint CO(2-1) structures. However, these artifacts are below the level of emission that we present in this paper and will be mitigated once the program's sensitive, high-fidelity data of the 12m array are added.

A standard calibration strategy was adopted by the ALMA observatory. The ACA data were calibrated using the ALMA calibration pipeline version 2023.1.0.124 \citep{hun23}. We imaged these pipeline-calibrated data using the Common Astronomy Sofware Applications (CASA) version 6.6.0.20 \citep{casa22}. After combining the two contiguous spectral windows around the CO(2-1) line using the {\it mstransform} task in CASA, we subtracted the continuum in the ({\em u},{\em v})-domain using CASA's {\it uvcontsub} task by fitting a straight line to the line-free channels. We then imaged the line-data with task {\it tclean}, using the mosaic gridder with a robust (Briggs) weighting, setting a robustness parameter of +0.5 \citep{bri95}, channel width of 5 \kms, and pixel-size of 1 arcsec. We also cleaned the line signal in each channel using a multi-scale deconvolution with scale levels set to 0, 6, 14, and 30 pixels until an absolute value for the threshold of 32 mJy\,beam$^{-1}$ was reached. This resulted in a line-data cube with a synthesized beam of 8.0$^{\prime\prime}$\,$\times$\,7.0$^{\prime\prime}$ (PA\,=\,-45.2$^{\circ}$) and a root-mean-square (rms) noise of 8.5 mJy\,beam$^{-1}$ per channel. A correction for the primary beam response was applied to this mosaic image to recover the true flux densities towards the edges of the field.

To construct moment maps, we first created a mask to extract the line signal. For this, we re-ran the above imaging steps, but this time using natural weighting and not performing the primary beam correction. We then wrote all CASA image products into Flexible Image Transport System \citep[FITS;][]{hanisch01,pen10} files and loaded them into the MIRIAD software \citep{sault95}. We created a mask from the naturally weighted image cube by replacing all negative signal with `0' values, applying a Hanning smooth, followed by a spatial smooth by convolving the data with a circular Gaussian beam of 5$^{\prime\prime}$, and finally applying a second Hanning smooth. This resulted in a smoothed data set with a synthesized beam of 9.5$^{\prime\prime}$\,$\times$\,8.7$^{\prime\prime}$ (PA\,=\,-44.6$^{\circ}$) that was used for creating the mask. All signal above 18 mJy\,beam$^{-1}$, which corresponds to 4$\sigma$ had we not replaced the negative signal with `0' values, was used as a mask to extract the signal in the full-resolution, primary-beam-corrected image cube. This resulted in a ``primary masked'' image cube, from which we made a preliminary total intensity (moment-0) image by summing all the signal. Some low-level noise remained visible in this moment map at the edges of the mosaic, therefore we re-did the masking and moment-0 extraction on the full resolution image cube but with no primary-beam correction applied, and then created a mask from this by setting a limit of 0.4 Jy\,beam$^{-1}$\,$\times$\,\kms. After pulling the ``primary masked'' image cube through this second mask, we created the final total-intensity (moment-0) and velocity (moment-1) maps.

\subsection{Combined Array for Research in Millimeter-wave Astronomy (CARMA)}
\label{sec:carma}

CARMA observations of the CO(1-0) line ($\nu_{\rm rest}$\,=\,115.27\,GHz) were made with three separate tracks on 2010 Aug 8, 9 and 10 totalling 10.50 hrs of observing time (project c0593). Unlike the ACA mosaic observations, the CARMA observations comprised of a single field center at RA(J2000) = 22h~35m~59.8s and Dec(J2000) =  33d~58m~16.6s. The primary beam response thus drops rapidly away from this pointing center, resulting in strong attenuation of the detected line emission outside the full width at half the maximum intensity (FWHM) of the primary beam at 73$^{\prime\prime}$. This includes the outer parts of NGC~7319, the Northern region near SQ-A, and much of the outer parts on the West. Moreover, because CARMA is a heterogeneous array, consisting of both 6m and 10m telescopes, the overall primary beam response in these regions is rather complex. Details of the CARMA observations have been given elsewhere. For example, \citet[][see their Fig.\,14]{appleton17} presented a preliminary total intensity map of the CO(1-0) emission, and more information about the data reduction was presented in \citet{appleton23}, where spectra were extracted and compared with ACA CO(2-1) data.

A total intensity (moment-0) map of the CARMA CO(1-0) data was constructed using the same method as for the ACA data. Initially a cube was created with a synthesized beam of 4.1$^{\prime\prime}$\,$\times$\,3.3$^{\prime\prime}$, as described by \citet{appleton23}. In order to better compare the CARMA and ACA data, we further smoothed the CARMA data to 8$^{\prime\prime}$\,$\times$\,7$^{\prime\prime}$ (PA\,=\,-45$^{\circ}$), with a channel width of 20 \kms. From this, we created a mask by first masking out negative signal, then Hanning smoothing the cube, convolving the signal with a 5$^{\prime\prime}$ Gaussian to a resolution of 9.4$^{\prime\prime}$\,$\times$\,8.6$^{\prime\prime}$, and applying a second Hanning smooth. To create the final mask, we then set a cutoff of 8 mJy\,beam$^{-1}$, which corresponds to 4$\sigma$ had we not replaced the negative signal with `0' values. The 8$^{\prime\prime}$\,$\times$\,7$^{\prime\prime}$ data cube with 20 \kms\ channels was then pulled through this mask, and a moment-0 map was created by summing all the unmasked signal. 

\begin{figure*}
\centering \includegraphics[width=\textwidth]{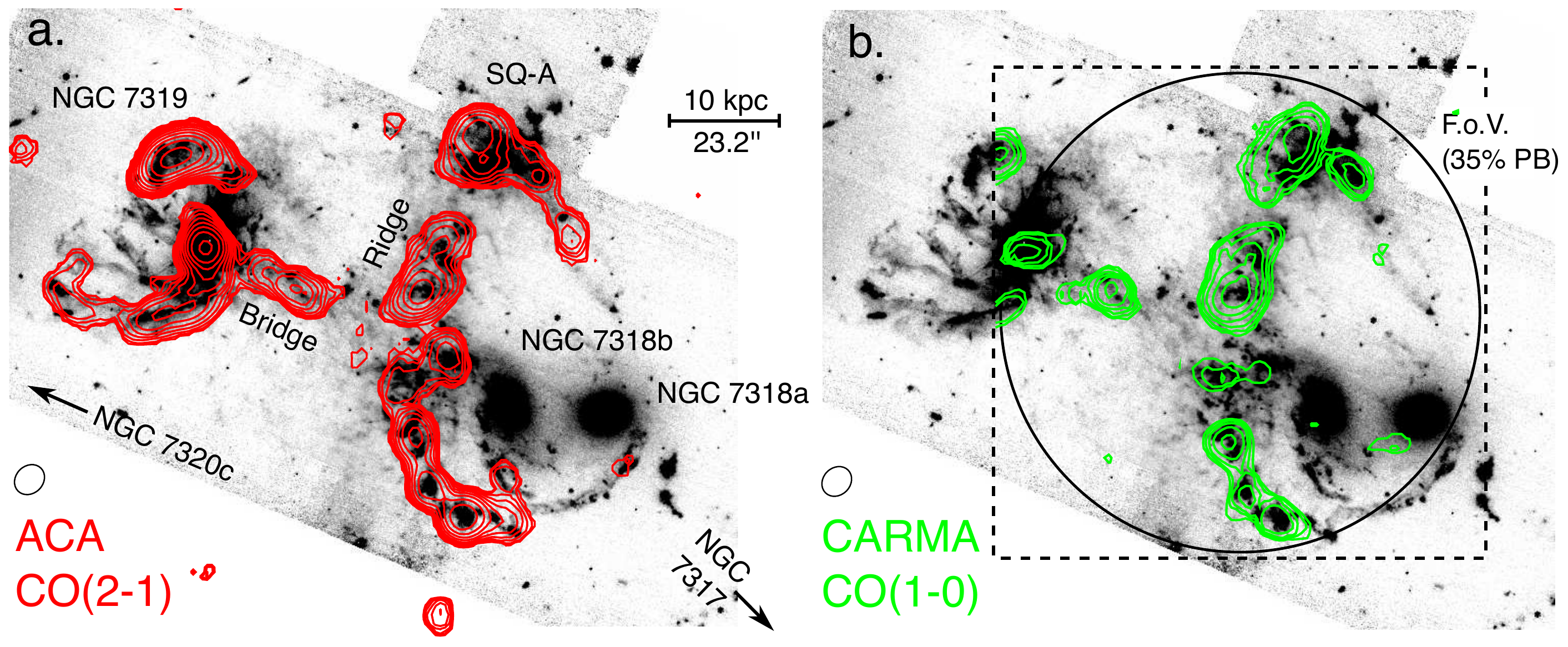}
\caption{CO emission in \SQ. Left (a): Red contours of the integrated CO (2-1) emission obtained with the ACA (resolution 8$^{\prime\prime}$\,$\times$\,7$^{\prime\prime}$), superimposed on a 10$\mu$m inverted gray-scale image obtained with JWST MIRI in the F1000W band \citep{pontoppidan22,doijwst}. Contours levels start at 0.41 Jy\,beam$^{-1}$\,\kms\ and increase by a factor $\sqrt{2}$. As shown by \citet{appleton23}, the background image is dominated by warm molecular hydrogen emission in the pure-rotational line 0-0\,S(3)\,9.66$\mu$m. The ACA mosaic covers the full region of the {\it JWST} background image shown here. Right (b): Green contours of the integrated CO (1-0) emission obtained with CARMA (smoothed to 8$^{\prime\prime}$\,$\times$\,7$^{\prime\prime}$), superimposed on the same MIRI image. Contour levels are the same as for the CO(2-1), but for visualization purposes the flux densities are not corrected for the effects of the complex primary beam attenuation. The black circle shows the 35$\%$-level of primary beam, while the dotted square marks the size of the masked image cube.}
\label{fig:acacarma}
\end{figure*}

\section{Results}
\label{sec:results}

\begin{figure*}
\centering \includegraphics[width=\textwidth]{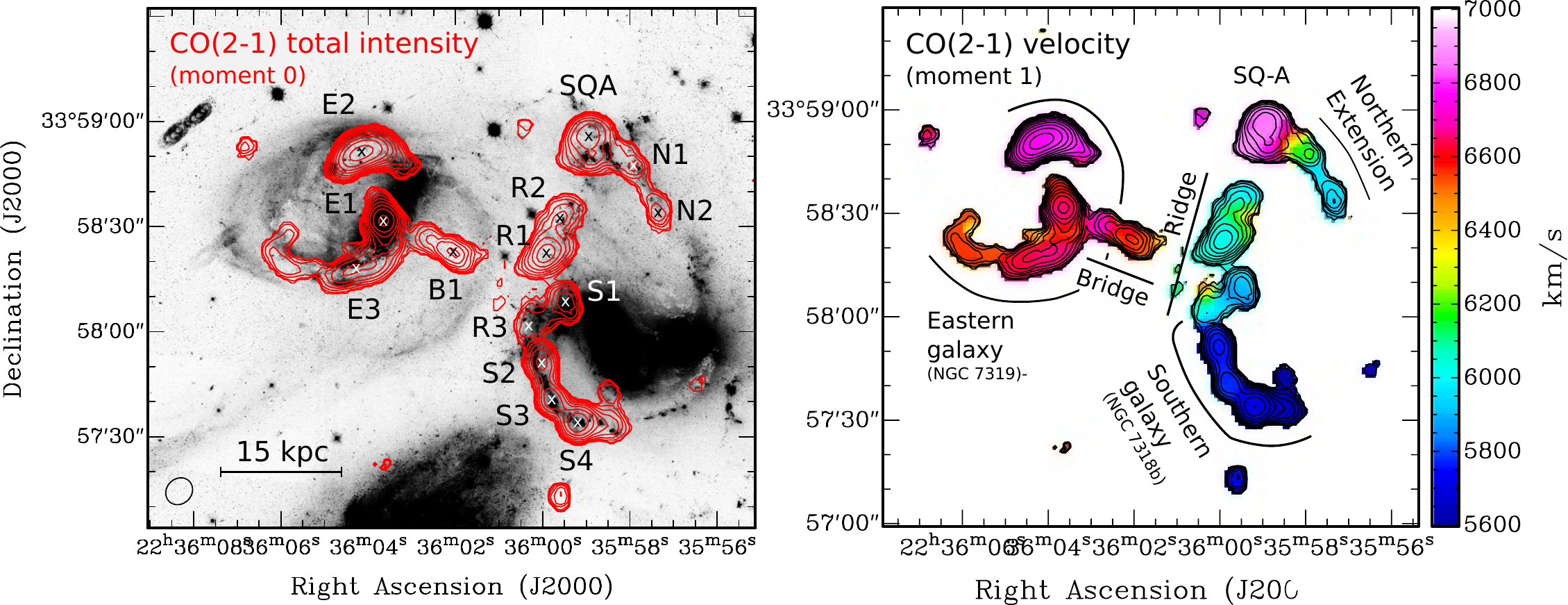}
\caption{Left: Total intensity (moment 0) map of the CO(2-1) emission obtained with the ACA. Contour levels start at 0.4 \jybmkms\ and increase by a factor $\sqrt{2}$. The background is an archival image taken with the {\it Hubble Space Telescope} ({\it HST}) Wide Field Camera 3 (WFC3) using the B-band F438W filter \citep{fedotov11}. The different regions that we describe in this paper are annotated, while the small crosses indicate the positions against which CO(2-1) and CO(1-0) spectra were extracted. Right: CO(2-1) velocity (moment 1) map, based on the optical heliocentric definition. Overlaid are the contours of the CO(2-1) total intensity. The different structures that we describe in this paper are also annotated.}
\label{fig:mapregions}
\end{figure*}

\begin{figure*}
\centering \includegraphics[width=\textwidth]{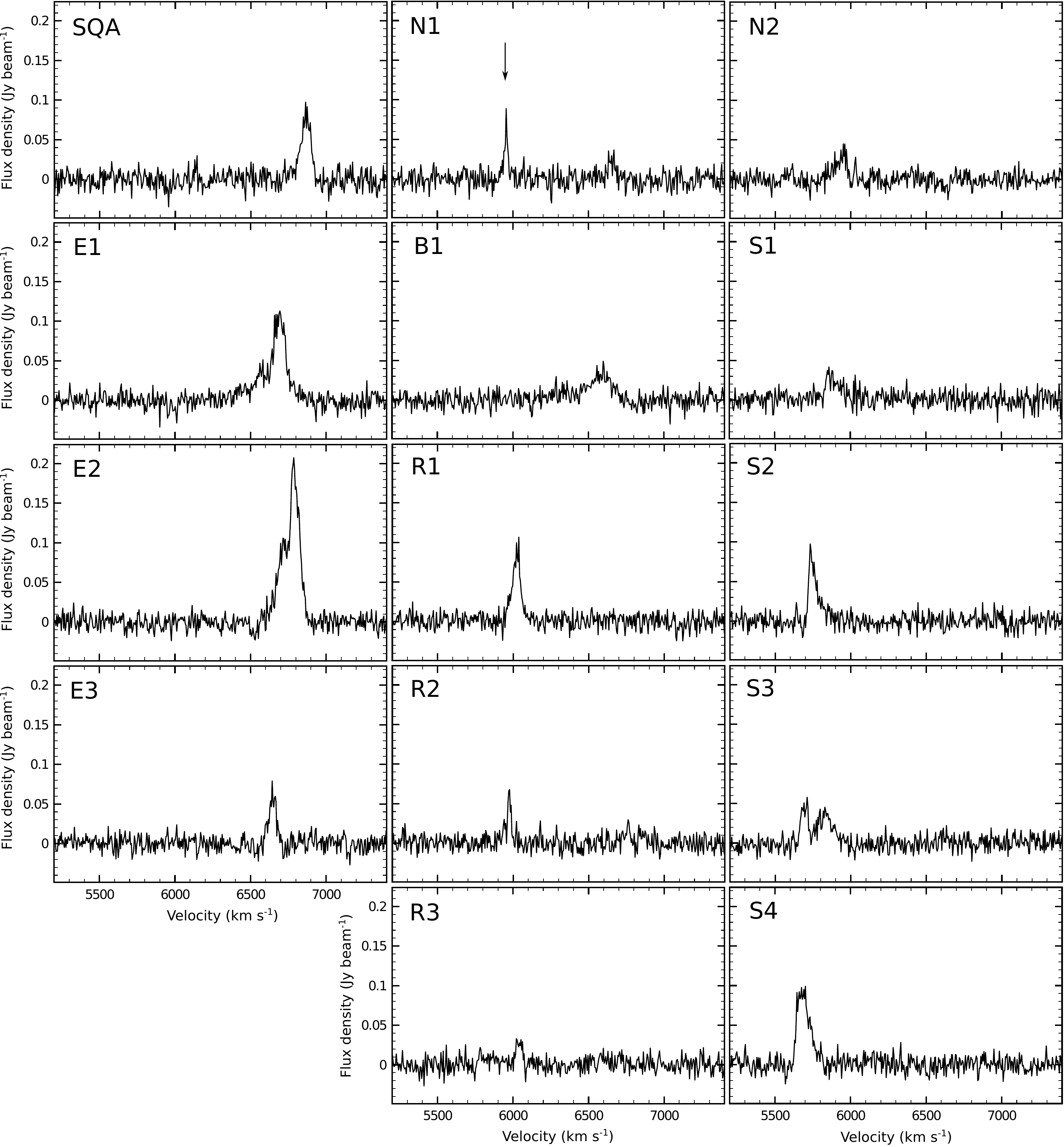}
\caption{CO(2-1) spectra from the ACA data in the regions indicated in Fig.\,\ref{fig:mapregions}. For region N1, the arrow points to the CO(2-1) component shown in Table \ref{tab:results}, while the weak secondary component around 6650 \kms\ will be described later in Fig.\,\ref{fig:SQA}.}
\label{fig:spec1}
\end{figure*}

Figure \ref{fig:acacarma} shows the total intensity images of CO(2-1) from the ACA data (left) and CO(1-0) from the CARMA data (right), superposed onto a {\it James Webb Space Telescope} ({\it JWST}) image made with the F1000W filter of the Mid-Infrared Instrument (MIRI) \citep{pontoppidan22,doijwst}. The ACA map shows that we detect bright CO(2-1) emission of cold molecular gas in the intra-group medium, in particular along the shock-front that stretches North to South in between the merging galaxies (the `ridge'), in a perpendicular structure between the ridge and NGC\,7319 (the `bridge'), and in the Northern starforming region SQ-A. Furthermore, we detect CO(2-1) in galaxy NGC\,7319 and along the Southern spiral arm of galaxy NGC\,7318b (i.e., the arm that faces the ridge).

For the most part, the CO(2-1) emission follows the same approximate distribution as the bright emission in the 10$\mu$m MIRI image. The 10$\mu$m emission has been shown to be dominated by warm (T\,$>$\,100\,K) molecular gas in the 0-0\,S(3) pure-rotational line \citep{appleton23}, except in regions of strong continuum emission. This suggests that the cold and warm H$_{2}$ are co-spatial on large scales. The only notable exception is the South-Western part of \SQ, where a second arm that is bright at 10$\mu$m overlays with the central parts of the NGC\,7318a/b pair to form the `smiley face' visible in Fig.\,\ref{fig:acacarma}. We do not detect significant CO(2-1) in either this second arm or the central regions of NGC\,7318a/b, despite the fact that the number of young star clusters in this region is as high as elsewhere in \SQ\ \citep{fedotov15}.

The CO(2-1) and CO(1-0) emission follow roughly the same distributions. In particular the brightest knots of emission along the main shocked ridge, the Southern arm of NGC\,7318b, the bridge, and SQ-A are well represented in both maps. As explained in Sect.\,\ref{sec:carma}, the CARMA moment-0 map (Fig.\,\ref{fig:acacarma}b) is much less sensitive to emission in NGC 7319 and only the nucleus is portrayed in full in this image. Because CARMA is a heterogeneous array with telescope dishes of different size (Sect.\,\ref{sec:carma}), we deem corrections of the primary beam response of the CO(1-0) data to be unreliable beyond the radius where the strength of the signal drops off to below 35$\%$ compared to the phase center (large circle in Fig.\,\ref{fig:acacarma}, right). Even though we detect CO(1-0) in regions further out, such as regions E2 and E3, we leave these out of the analysis presented in this paper.

Figure \ref{fig:mapregions} (left) shows again the ACA total intensity image of CO(2-1) overlaid onto a B-band image taken with the {\it Hubble Space Telescope} ({\it HST}), while Fig. \ref{fig:mapregions} (right) shows a velocity (moment-1) map of the CO(2-1). Based on both the spatial distribution and kinematics we identify six global structures, with some subdivided into several regions, which are all annotated in Fig.\,\ref{fig:mapregions}. Individual spectra were extracted against the peak emissions in each region. The CO(2-1) spectra are shown in Fig.\,\ref{fig:spec1}. The spectra were used to derive the observed CO properties in each region, which are summarized in Table\,\ref{tab:results}. The full spectral analysis can be found in Appendix \ref{sec:app_spectral}. As part of this spectral analysis, we also estimate the line luminosity of CO(2-1) and CO(1-0) \citep{solomon05}, and luminosity ratio r$_{2-1}$\,=\,$L^{\prime}_{\rm CO(2-1)}$/$L^{\prime}_{\rm CO(1-0)}$. In the remainder of this Section, we will described the observational results for the individual structures in \SQ.

\begin{deluxetable*}{lcccccccc}
\tablecaption{Table summarizing the ACA and CARMA properties in the various regions. More details are given in Appendix \ref{sec:app_spectral}.}
\tablehead{
\colhead{Region} & \colhead{R.A.} & \colhead{Dec.} &  \colhead{$v_{\rm opt}$} & \colhead{$\int_{v} S_{\rm CO(2-1)} \delta {v}$} & \colhead{$\int_{v} S_{\rm CO(1-0)} \delta {v}$} & \colhead{$L^{\prime}_{\rm CO(2-1)}$\,$^{\dagger}$} & \colhead{$L^{\prime}_{\rm CO(1-0)}$} & \colhead{r$_{21}$} \\
\colhead{} & \colhead{J2000} & \colhead{J2000} & (\kms) & \multicolumn{2}{c}{(\jybmkms)} & \multicolumn{2}{c}{($\times$10$^{7}$~\Kkmspc)} & \colhead{} \\
\colhead{} & \colhead{} & \colhead{} & \colhead{} & \colhead{} & \colhead{} & \colhead{} & \colhead{$\equiv$\,M$_{\rm H_2}$/$\alpha_{\rm CO}$\,$^{\ddagger}$} & \colhead{} 
}
\startdata
R1 & 22$^{\rm h}$35$^{\rm m}$59.91$^{\rm s}$ & 33$^{\circ}$58$^{\prime}22.13^{\prime\prime}$ & 6022 & 5.73\,$\pm$\,0.25 & 2.67\,$\pm$\,0.18 & 3.0\,$\pm$\,0.1 & 5.7\,$\pm$\,0.4 & 0.54\,$\pm$\,0.06 \\
R2 & 22$^{\rm h}$35$^{\rm m}$59.58$^{\rm s}$ & 33$^{\circ}$58$^{\prime}$32.13$^{\prime\prime}$ & 5974 & 1.76\,$\pm$\,0.18 & 1.07\,$\pm$\,0.13 & 0.9\,$\pm$\,0.1 & 2.3\,$\pm$\,0.3 & 0.41\,$\pm$\,0.09 \\
R3 & 22$^{\rm h}$36$^{\rm m}$00.31$^{\rm s}$ & 33$^{\circ}$58$^{\prime}$01.13$^{\prime\prime}$ & 6039 & 1.45\,$\pm$\,0.18 & 0.67\,$\pm$\,0.14 & 0.8\,$\pm$\,0.1 & 1.4\,$\pm$\,0.3 & 0.55\,$\pm$\,0.19 \\
B1 & 22$^{\rm h}$36$^{\rm m}$02.08$^{\rm s}$ & 33$^{\circ}$58$^{\prime}$22.63$^{\prime\prime}$ & 6575 & 6.00\,$\pm$\,0.40 & 4.12\,$\pm$\,0.43 & 3.2\,$\pm$\,0.2 & 8.7\,$\pm$\,0.9 & 0.36\,$\pm$\,0.06 \\
SQ-A & 22$^{\rm h}$35$^{\rm m}$58.94$^{\rm s}$ & 33$^{\circ}$58$^{\prime}$56.13$^{\prime\prime}$ & 6864 & 6.52\,$\pm$\,0.32 & 4.75\,$\pm$\,0.41 & 3.4\,$\pm$\,0.2 & 10.1\,$\pm$\,0.9 & 0.34\,$\pm$\,0.05 \\
N1$^{\S}$ & 22$^{\rm h}$35$^{\rm m}$57.93$^{\rm s}$ & 33$^{\circ}$58$^{\prime}$47.13$^{\prime\prime}$ & 5956 & 1.79\,$\pm$\,0.19 & 1.54\,$\pm$\,0.24 & 0.9\,$\pm$\,0.1 & 3.3\,$\pm$\,0.5 & 0.29\,$\pm$\,0.08 \\
N2 & 22$^{\rm h}$35$^{\rm m}$57.33$^{\rm s}$ & 33$^{\circ}$58$^{\prime}$33.63$^{\prime\prime}$ & 5935 & 3.05\,$\pm$\,0.36 & 1.49\,$\pm$\,0.36 & 1.6\,$\pm$\,0.2 & 3.2\,$\pm$\,0.7 & 0.51\,$\pm$\,0.18 \\
S1 & 22$^{\rm h}$35$^{\rm m}$59.46$^{\rm s}$ & 33$^{\circ}$58$^{\prime}$08.63$^{\prime\prime}$ & 5882 & 3.42\,$\pm$\,0.34 & 0.99\,$\pm$\,0.21 & 1.8\,$\pm$\,0.2 & 2.1\,$\pm$\,0.4 & 0.86\,$\pm$\,0.27 \\
S2 & 22$^{\rm h}$36$^{\rm m}$00.03$^{\rm s}$ & 33$^{\circ}$57$^{\prime}$51.13$^{\prime\prime}$ & 5739 & 5.93\,$\pm$\,0.82 & 2.93\,$\pm$\,0.38 & 3.1\,$\pm$\,0.4 & 6.2\,$\pm$\,0.8 & 0.51\,$\pm$\,0.14 \\
S3 & 22$^{\rm h}$35$^{\rm m}$59.78$^{\rm s}$ & 33$^{\circ}$57$^{\prime}$41.13$^{\prime\prime}$ & 5762$^{*}$ & 7.17\,$\pm$\,0.38 & 3.81\,$\pm$\,0.65 & 3.8\,$\pm$\,0.2 & 8.1\,$\pm$\,1.4 & 0.47\,$\pm$\,0.11 \\
S4 & 22$^{\rm h}$35$^{\rm m}$59.18$^{\rm s}$ & 33$^{\circ}$57$^{\prime}$34.13$^{\prime\prime}$ & 5687$^{*}$ & 9.43\,$\pm$\,2.11 & 3.65\,$\pm$\,0.69 & 5.0\,$\pm$\,0.1 & 7.7\,$\pm$\,1.5 & 0.65\,$\pm$\,0.27 \\
E1 & 22$^{\rm h}$36$^{\rm m}$03.64$^{\rm s}$ & 33$^{\circ}$58$^{\prime}$31.63$^{\prime\prime}$ & 6694 & 16.60\,$\pm$\,0.85 & 5.74\,$\pm$\,1.13 & 8.8\,$\pm$\,0.4 & 12.1\,$\pm$\,2.4 & 0.72\,$\pm$\,0.18 \\
E2 & 22$^{\rm h}$36$^{\rm m}$04.17$^{\rm s}$ & 33$^{\circ}$58$^{\prime}$51.13$^{\prime\prime}$ & 6794 & 22.00\,$\pm$\,0.96 & - & 11.6\,$\pm$\,0.5 & (23.2\,$\pm$\,1.0)$^{**}$ & - \\
E3 & 22$^{\rm h}$36$^{\rm m}$04.33$^{\rm s}$ & 33$^{\circ}$58$^{\prime}$18.13$^{\prime\prime}$ & 6646 & 3.71\,$\pm$\,0.22 & - & 2.0\,$\pm$\,0.1 & (3.9\,$\pm$\,0.2)$^{**}$ & - \\
\enddata
\tablecomments{$^{\dagger}$ $L^{\prime}_{\rm CO}$ was calculated following Equation 3 of \citet{solomon05}, and assuming redshift $z$\,=\,0.0215 and luminosity distance $D_{\rm L}$\,=\,94 Mpc.\\
$^{\ddagger}$ Mass is in units of $M_{\odot}$; $\alpha_{\rm CO}$\,$\equiv$\,$M_{\rm H_2}$/$L^{\prime}_{\rm CO(1-0)}$ is the CO conversion factor in units of $M_{\odot}$/(\Kkmspc) \citep[e.g.,][]{bolatto13}. \\
$^{\S}$ Spectral component marked with the arrow in Fig.\,\ref{fig:spec1}.\\
$^{*}$ Average velocity of the two overlapping components (see Appendix \ref{sec:app_spectral}).\\
$^{**}$ Regions E2 and E3 lie too far beyond the FWHM of the primary beam to measure accurate CO(1-0) flux densities. We based these estimates on $S_{\rm CO(2-1)}$ and assume r$_{21}$\,=\,0.5, which is the weighted average value among the other regions (Sect.\,\ref{sec:mass}).
}
\label{tab:results}
\end{deluxetable*}

\subsection{Ridge (R).}
\label{sec:ridge}

The `ridge' is the main shock-front that crosses \SQ\ (Sect.\,\ref{sec:intro}). We detect CO(2-1) and CO(1-0) along the ridge, with an average gas excitation of r$_{21}$\,=\,0.5\,$\pm$\,0.2. We classify region R3 as part of the ridge, because its velocity is closer to that of the CO(2-1) in regions R1 and R2 than to the velocity seen across the arm of NGC\,7318b (S1$-$S4); see Fig.\,\ref{fig:mapregions} (right).

\subsection{Bridge (B).} 
\label{sec:bridge}

\begin{figure}
\centering \includegraphics[width=0.95\columnwidth]{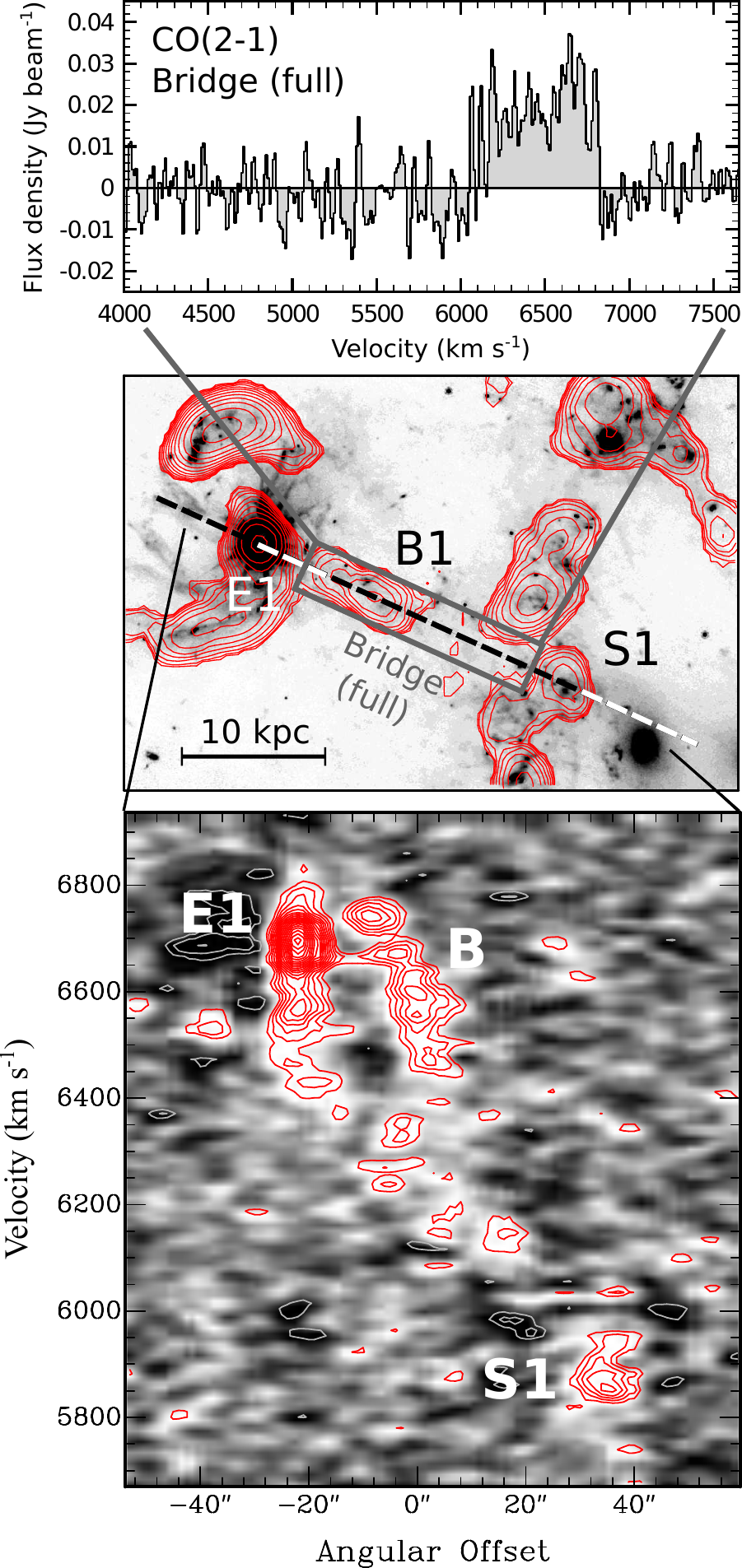}
\caption{CO(2-1) emission observed with the ACA across the bridge. To visualize the faint signal, the data were binned by two channels to a channel width of 10 \kms\ and subsequently Hanning smoothed to an effective velocity resolution of 20 \kms. Top: CO(2-1) spectrum along the bridge, obtained by integrating the signal within the gray box in the middle panel (dashed line). Middle: CO(2-1) total intensity image from Fig.\,\ref{fig:mapregions} with annotated a 1-dimensional pseudo-slit, which effectively captures signal across one synthesized beam in the direction parallel to the slit. Bottom: position-velocity map of CO(2-1) along the pseudo-slit shown in the middle panel. The pseudo-slit covers the full length of the bridge, as well as regions E1 and S1. Red contours start at 2$\sigma$ and increase with 1$\sigma$, with $\sigma$\,=\,5 mJy\,beam$^{-1}$ the rms noise (the same for negative signal in gray contours).}
\label{fig:bridge}
\end{figure}

The `bridge' is an elongated structure of CO emission that stretches in between NGC\,7319 and NGC\,7318b, along a direction that is perpendicular to the ridge. The Eastern end connects to the central region of NGC\,7319. In the CO(2-1) total intensity image of Fig.\,\ref{fig:acacarma}, the Western end appears to stops mid-way towards NGC\,7318b, approximately 1$-$2 beam-sizes away from the ridge. However, because our ACA data are sensitive to emission at low surface brightness, the position-velocity map in Fig.\,\ref{fig:bridge} shows faint emission stretching all the way to the ridge, across $\sim$10 kpc and covering $\sim$700 \kms.  

The CO in the bright central part of the bridge shows a kinematic gradient in spatial direction (Fig.\,\ref{fig:bridge}). At the peak of the emission, the CO(2-1) line has a FWHM\,$\sim$\,184 \kms. This is significantly wider than in other regions (Table \ref{tab:fitting}), with the exception of a broad, blueshifted component in the center of NGC\,7319 (region E1), which is also visible in the position-velocity map of Fig.\,\ref{fig:bridge}. Midway between NGC\,7319 and NGC\,7318b, it appears that the bridge and the blueshifted component from region E1 merge into a single structure that connects to the ridge. The gas excitation of r$_{21}$\,=\,0.36\,$\pm$\,0.06 in region B1 appears to be lower than the gas excitation along the ridge.

\subsection{NGC\,7318a/b (S).}
\label{sec:7318}

NGC\,7318a and NGC\,7318b are two galaxies towards the South-West of the group. 
NGC\,7318b contains a prominent arm on the East, where it coincides with the ridge. We detect CO(2-1) emission along this arm, with a bright knot of CO(2-1) emission North-East of the nucleus (S1 in Fig.\,\ref{fig:mapregions}), and additional CO(2-1) along the arm as it extends to the South (S2\,$-$\,S4). No CO(2-1) is detected from the inner regions of NGC\,7318a/b, and also the Western side of this galaxy pair is notably devoid of CO (Sect.\,\ref{sec:results}). 

The distribution of CO in NGC\,7318a/b is very similar to the distribution of H$\alpha$-emitting gas found by \citet{rodriguez14}. The lack of detectable gas in the main body of these two galaxies suggests that much of the gas has been stripped into the intergalactic medium by collisions between the galaxies in the past. This is supported by the detection of faint, likely tidally stripped, \HI\ in the far outer reaches of the \SQ\ group by \citet{xu22}.

\subsection{NGC\,7319 (E).} 
\label{sec:7319}

\begin{figure*}
\centering \includegraphics[width=\textwidth]{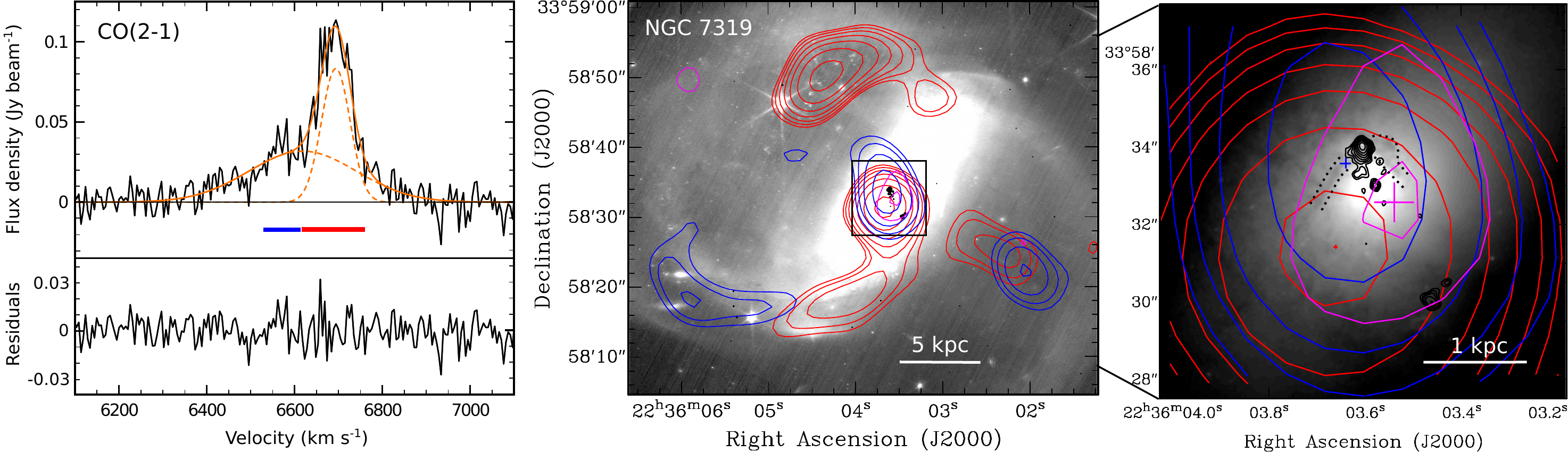}
\caption{CO(2-1) emission observed with the ACA in NGC\,7319. Left: Spectrum of the central region E1, with overlaid a model (orange line) that consist of two Gaussian component (dashed orange lines). Residuals after subtracting the model from the spectrum are given at the bottom. The blue and red bar indicate the velocity ranges across which we made total intensity images shown in the other two panels (blue: 6529$-$6610 \kms; red: 6610$-$6768 \kms). Middle: {\it JWST} NIRCam/F090W image \citep{pontoppidan22,doijwst} with overlaid in red and blue the contours of the CO(2-1) total intensity emission across the velocity ranges indicated on the left panel. Contour levels start at 5$\sigma$ and increase by factor $\sqrt{2}$, with $\sigma$\,=\,0.24 and 0.17 \jybmkms\ for the red and blue map, respectively. The magenta contours show the 240\,GHz continuum of our ACA data. Contours are drawn at 3 and 4$\sigma$, with $\sigma$\,=\,0.36 mJy\,beam$^{-1}$. The black contours show the Merlin 1.6\,GHz radio continuum \citep{xanthopoulos04}. Contour levels start at 0.2 mJy\,beam$^{-1}$ and increase with a factor $\sqrt{2}$. Right: Zoom-in of the central region shown in the middle plot. The dotted lines highlight a dark lane that crosses in projection the Northern radio lobe. The crosses mark the relative uncertainty in the location of the peak emission of CO and dust ($\delta \theta_{\rm rms}$), calculated following $\delta \theta_{\rm rms}$\,$\sim$\,$\frac{1}{2} \langle \Theta_{\rm beam} \rangle$(S/N)$^{-1}$, with $\Theta_{\rm beam}$ the ACA beam size and S/N the signal-to-noise of the peak emission \citep[following][]{papadopoulos2008}.}
\label{fig:AGN}
\end{figure*}

NGC\,7319 is the galaxy towards the East. It contains more than 40$\%$ of the total CO(2-1) luminosity that our ACA data recover in \SQ\ (Table \ref{tab:results}), despite the fact the galaxy was fully stripped of its \HI\ gas \citep{shostak84}. 

Our ACA data show unresolved 1.3mm continuum emission with $S_{\rm 240\,GHz}$\,=\,1.5 mJy (4$\sigma$) at the location of the Seyfert-2 AGN (Fig.\,\ref{fig:AGN}). The CO(2-1) spectrum at this location shows a prominent blue wing, with a broad FWHM\,=\,289\,$\pm$\,18\,\kms\ (Table\,\ref{tab:fitting}). The emission-line peak of this broad component is blueshifted by 75 \kms\ with respect to the peak of the narrow component (Table\,\ref{tab:fitting}). The latter has a velocity of $v_{\rm narrow}$\,=\,6694\,$\pm$\,3 \kms, which is consistent with the systemic velocity of 6740\,$\pm$\,50 \kms\ derived from Mg\,{\it b} absorption by \citet{aoki96}. Figure \ref{fig:AGN} (right panel) shows that the CO(2-1) emission in the blue wing is offset by 1.5$^{\prime\prime}$ ($\sim$0.7\,kpc) from the peak of the 1.3mm continuum emission and coincides with the bright NE hot-spot of the radio source, which is also the location of a dark lane in the {\it JWST} F090W imaging \citep[][]{pereira22}. Therefore, it is likely that the blueshifted component either represents an outflow of molecular gas driven by the radio source, or it is turbulent molecular gas along an inner dust lane. We will discuss this in more detail in Sect.\,\ref{sec:natureoutflow}.

For the narrow CO(2-1) component in the center of NGC\,7319, the peak of the emission is offset by 1.8$^{\prime\prime}$ ($\sim$0.8\,kpc) SE of the dust continuum. This peak emission likely includes bright CO at the base of the southern arm that stretches in the direction of region E3 in Fig.\,\ref{fig:mapregions}. Based on the spectral decomposition performed in Appendix \ref{sec:app_spectral} (Table \ref{tab:fitting}), the molecular gas of the narrow component shows an excitation consistent with unity (r$_{21}$\,=\,1.36\,$\pm$\,0.64), albeit with a large uncertainty. The excitation in the blue wing of the profile appears to be substantially lower (r$_{21}$\,=\,0.54\,$\pm$\,0.16). 

\begin{figure*}
\centering \includegraphics[width=\textwidth]{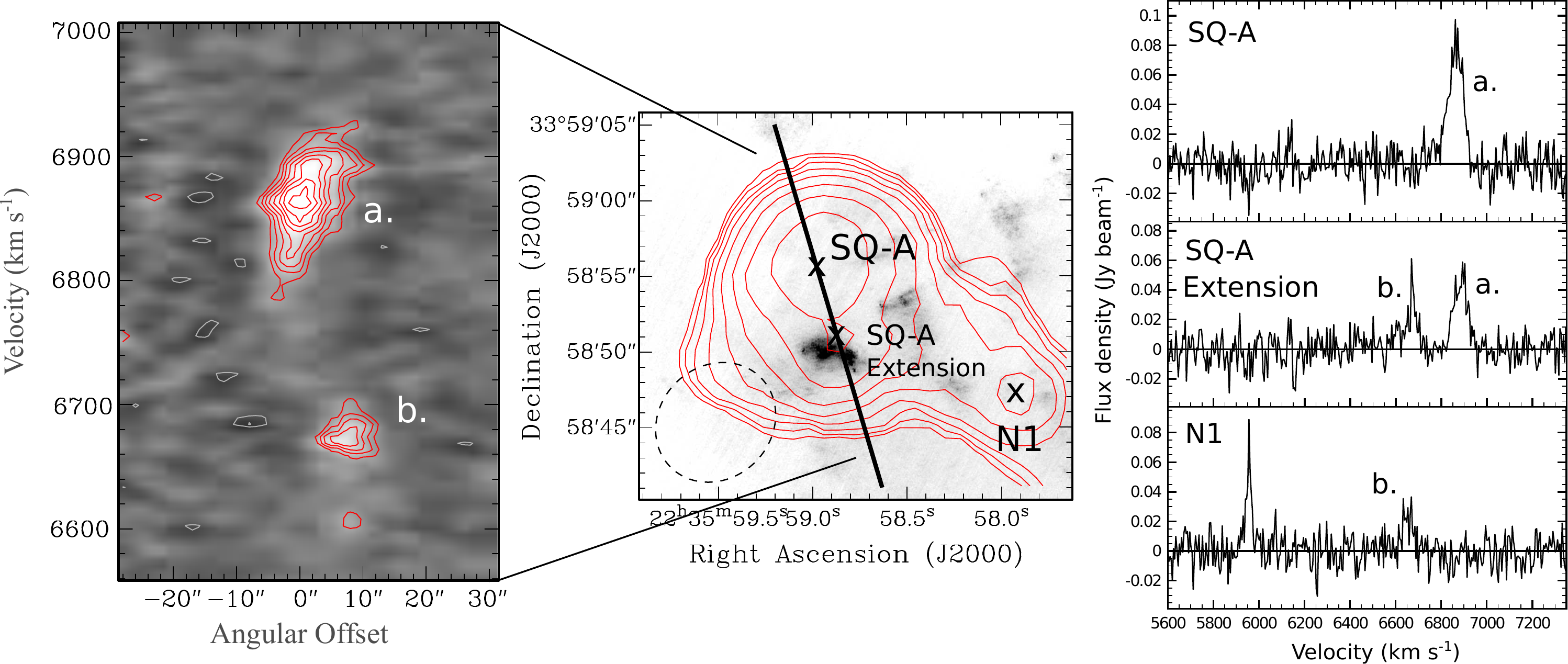}
\caption{CO(2-1) emission observed with the ACA in SQ-A. Left: Position-velocity plot taken along the solid line in the middle panel, after Hanning smoothing the data. Contour levels start at 2.5$\sigma$ and increase with 1$\sigma$, with $\sigma$\,=\,0.08 mJy\,beam$^{-1}$. Component {\sl a} is CO(2-1) emission at the velocity of the the CO peak in SQ-A, while component {\sl b} is a secondary velocity component that peaks towards the South, in a region that we label `SQ-A Extension'. Middle: contours of the CO(2-1) total intensity overlaid onto the {\it JWST}/F090W image of SQ-A, highlighting a bright IR source near kinematic component {\sl b} in SQ-A Extension. Contour levels are the same as in Fig.\,\ref{fig:mapregions}. Right: CO(2-1) spectra of regions SQ-A, SQ-A Extension, and N1, extracted against the position marked by the crosses in the middle panel}.
\label{fig:SQA}
\end{figure*}

Region E2, in the northern part of NGC~7319, has the brightest CO(2-1) emission found in Stephan's Quintet, and was detected by CARMA in CO(1-0) despite being far from the primary beam center (Fig.\,\ref{fig:acacarma}). Also here, the spectrum consists of two components (Fig.\,\ref{fig:spec1}). Both components have a FWHM\,$\sim$\,80 \kms, which is similar to the narrow component in region E1. The velocity separation between the two components is 90 \kms, and the systemic velocity of 6740 \kms\ \citep{aoki96} falls in the middle of the two components (see Table \ref{tab:fitting}). The {\it JWST} imaging of Fig.\,\ref{fig:AGN} (middle panel) shows indications for at least three diffuse, tail-like structures in the North-Eastern part of NGC\,7319. It is likely that in region E2 two of these tails overlap, causing the bright, multi-component CO(2-1) emission. At the same location in region E2, a major patch of star formation was seen in H$\alpha$ emission by \citet{xu99}. Region E3 marks the peak of the CO(2-1) emission along a southern arm of NGC\,7319. There are indications that the brightest CO(2-1) emission is spatially off-set from the brightest optical and near-IR emission in regions E2 and E3, which suggests that the CO aligns with dust lanes that stretch along the arms (Figs.\,\ref{fig:mapregions} and \ref{fig:AGN}). Both regions fall outside the 35$\%$-level of CARMA's primary beam, hence we do not derive r$_{\rm 21}$ values for these regions (see Sect.\,\ref{sec:results}).

\subsection{SQ-A (SQA).}
\label{sec:SQA}

SQ-A is the Northern region that is bright in the infrared and optical. This is a region of active star formation resolved into multiple components and structures with {\it JWST} \citep{appleton23}. Overall, SQ-A contains the coldest of the warm molecular gas detected by the {\it Spitzer} \citep{appleton17}. 

The CO in SQ-A shows the highest velocity of all the gas structures in \SQ\ ($v_{\rm opt}$\,$\sim$\,6864 \kms), although it is close to the CO velocities found across NGC\,7319. The gas excitation of r$_{21}$\,=\,0.34\,$\pm$\,0.05 is lower than for most of the other regions in \SQ. In Fig.\,\ref{fig:SQA} we show that this CO(2-1) emission peaks in a region that is devoid of strong IR emitters. However, a secondary, fainter CO(2-1) component is found $\sim$7$^{\prime\prime}$ (i.e., roughly one synthesized beam) towards the South at $v_{\rm opt}$\,$\sim$\,6650 \kms\ (component `b' in Fig.\,\ref{fig:SQA}). This secondary CO component has a bright IR counterpart in the {\it JWST} imaging, and is likely a region of star formation, but also displays a low r$_{21}$\,=\,0.36\,$\pm$\,0.09 (Table\,\ref{tab:fitting}). This component `b' is also faintly visible in the CO(2-1) spectrum of region N1, which suggests that this emission is spatially extended. The fact that both components (around 6650 and 6864 \kms) are clearly detected in CO(2-1) and CO(1-0) is consistent with single dish observations by \citet{lisenfeld02}, \citet{smith01}, and \citet{guillard12} (see also Sect.\,\ref{sec:excitation}). Using an `on-the-fly' technique for their single dish observations, \citet{yttergren21} also detected both components in CO(1-0), but they did not detect the stronger (6864 \kms) line in CO(2-1).

\subsection{Northern Extension (N).}
\label{sec:SQAext}

Extending southwest of SQ-A, the ACA and CARMA images reveal molecular gas at a lower velocity, which we will refer to as the `Northern Extension', based on its spatial location and stretched morphology. The Northern Extension has a distinct velocity around 5900 \kms, meaning that it is not kinematically associated with region SQ-A. Instead, the gas velocities in N1 and N2 are roughly the same as the gas in the ridge and offset by only $\sim$200 \kms\ from the gas along the Southern arm of NGC\,7318b. We will discuss this kinematic alignment with both the arm and the ridge in Sect. \ref{sec:kinematics}. The gas excitation along the Northern Extension, r$_{21}$\,$\la$\,0.3, is lower than across the ridge or the Southern arm of NGC\,7318b, and is in fact the lowest among all regions in \SQ.

\section{Discussion} 
\label{sec:discussion}

Our ACA and CARMA data provide a global overview of the cold molecular gas across the inner 70\,kpc of \SQ. In this Section, we will use the gas kinematics to create a 3-dimensional view of the group dynamics and discuss the properties of the cold gas, in particular the mass and excitation conditions.

\subsection{Global kinematics of the molecular gas}
\label{sec:kinematics}

\begin{figure*}
\centering 
\includegraphics[width=0.8\textwidth]{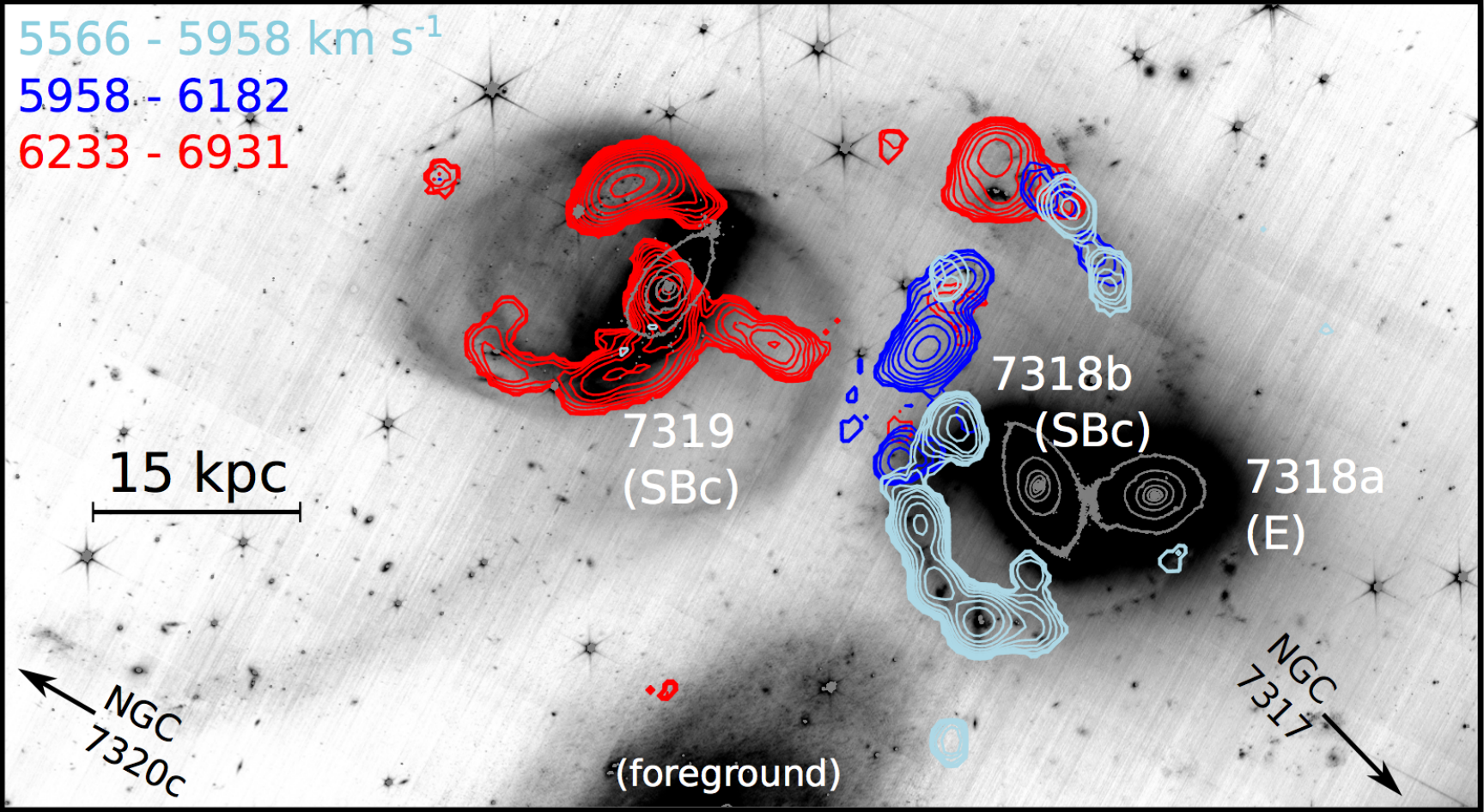}
\caption{Kinematic structures of the cold molecular gas in \SQ. Shown is a {\it JWST}/F090W image of \SQ\ \citep{pontoppidan22,doijwst}, with overlaid contours of total intensity CO(2-1) emission integrated across three velocity ranges, as indicated in the legend on the top-left. The CO contours start at 0.3 Jy\,beam$^{-1}$\,$\times$\,\kms\ and increase by a factor $\sqrt{2}$. The thin gray contours show the high surface brightness emission in the {\it JWST}/F090W image, starting at 0.1$\%$ of the peak intensity and increasing by a factor 2.}
\label{fig:jwstaca}
\end{figure*}

\begin{figure}
\centering 
\vspace{3mm}
\includegraphics[width=\columnwidth]{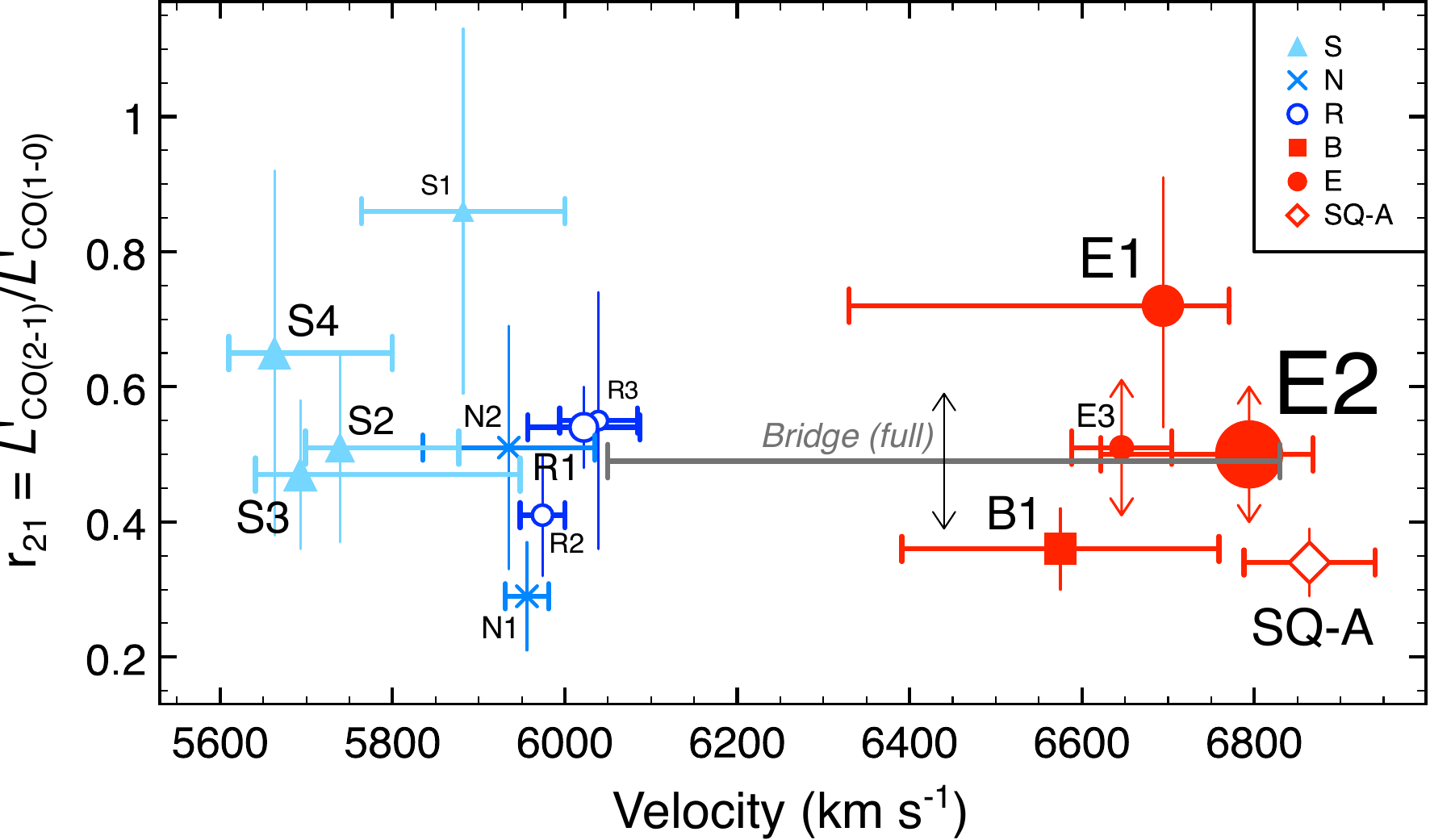}
\caption{Schematic overview of the observed properties of the various regions and structures, based on the Gaussian fitting presented in Tables \ref{tab:results} and \ref{tab:fitting}. Plotted on the horizontal axis is the central velocity of the CO(2-1) emission in each region listed in Table\,\ref{tab:results}. In case the line profile is fitted with two Gaussian components, the central velocity of the main narrow component is taken. The horizontal bar shows the full width at zero intensity (FWZI) of the line profile, which is twice the FWHM for each Gaussian component from Table\,\ref{tab:results} (except for the `full bridge', for which the FWZI is assumed to be 6050\,$-$\,6830 \kms\ from Fig.\,\ref{fig:bridge}). The vertical axis shows the CO line ratio (r$_{\rm 21}$), with the vertical bars the corresponding uncertainty (Table\,\ref{tab:results}). The different symbols mark the different regions, as per the legend in the top-right corner. The sizes of the symbols scale with the estimated H$_{2}$ mass, as per Table\,\ref{tab:results}. The color-coding highlights the low-, mid-, and high-velocity structures from Fig.\,\ref{fig:jwstaca}.}
\label{fig:propertyplot}
\end{figure}

In Fig.\,\ref{fig:jwstaca}, we compare the CO(2-1) distribution and kinematics with the deep {\it JWST}/F090W imaging to understand the relationship between the various regions. We identify three primary kinematical structures in our CO imaging of \SQ: a `low-$v$' component (5566-5958 \kms\ in Fig.\,\ref{fig:jwstaca}) associated with the `intruder' galaxy NGC7318b; a `mid-$v$' component (5958-6182 \kms) associated with the `ridge'; and a `high-$v$' component (6233-6931 \kms) associated with NGC\,7319, SQ-A, and the brightest part of the bridge. There is a notable gap in velocities between the high-$v$ and mid/low-$v$ components, with little cold gas in between. The only exception is the bridge, where faint 
CO(2-1) emission appears to stretch across $\sim$700 \kms, connecting the high-$v$ gas in NGC\,7319 with the mid- and low-$v$ gas in the ridge and NGC\,7318b across a distance of $\sim$10\,kpc (Fig.\,\ref{fig:bridge}). Figure \ref{fig:propertyplot} gives a schematic overview of the velocity information and line ratios derived from our CO data in the various regions of \SQ\ (based on Table\,\ref{tab:results}).

\subsubsection{Low/mid-$v$ components (NGC\,7318b/ridge)}

The mid-$v$ gas in the ridge has velocities that are close to that of the low-$v$ gas in NGC\,7318b, suggesting that the cold molecular gas that we detect along the ridge is closely related to the intruder galaxy NGC\,7318b. Figure \ref{fig:propertyplot} shows that the CO(2-1) lines are narrower across the ridge than in most of the other regions (also Table\,\ref{tab:fitting}), suggesting that the cold molecular gas that we detect is more settled along the ridge than elsewhere in \SQ. Contrary, the molecular gas along the spiral arm of NGC\,7318b has a redshifted wing or secondary redshifted component to the CO(2-1) profile in the regions S3\,$-$\,S4, and possible also in S1 (see Fig.\,\ref{fig:specS}). As can be seen in Fig.\,\ref{fig:propertyplot}, this red wing stretches to the velocities of the molecular gas in the ridge. We argue that this could resemble one of the scenarios described by \citet{guillard12}, where gas in the intruder galaxy is shocked and cooled when the galaxy falls into the group, and then decelerated to redder velocities. The gas in the ridge is likely either pre-existing gas, or decelerated material from the intruding NGC\,7318b. In case of the latter, this gas may be shocked by the intruder to experience a cooling time that is short compared to the free-fall time. 

To distinguish between star formation and shocks along the ridge, \citet{konstantopoulos14} studied optical emission lines among a sample of H$\alpha$-emitters. They found a mixture of narrow and broad lines, where the narrow lines have velocities of $v$\,$\sim$\,5800\,$-$\,6000 \kms\ and reflect H\,{\sc II} regions, while the broad lines have $v$\,$>$\,6000 \kms\ and are consistent with shocks that inhibit star formation in the IGM. Our CO(2-1) results only appear to trace counterparts to the narrow lines, with CO lines widths (FWHM\,$\sim$\,25\,$-$\,65 \kms) that are typical of those found in extra-giant galactic H\,{\sc II} regions \citep[e.g.,][]{roy86}. We do not find evidence in our CO data for much broader lines (several 100 \kms), which \citet{konstantopoulos14} associated with the shocked gas along the ridge, and which were previously also observed in the warm H$_{2}$ \citep{appleton06} and ionized gas \citep{iglesias12}. These broad shock tracers align in velocity with broad CO emission detected in single-dish data by \citet{guillard12}, but we find little to no evidence for this emission in our ACA data. In Sect.\,\ref{sec:scales} we will perform a more detailed comparison with previous single-dish CO data.

The regions that we previously defined as `Northern Extension' (N1 and N2 in Fig.\,\ref{fig:mapregions}) could either be the far end of a diffuse tail that stretches away from NGC\,7318b, or part of the same material that forms the ridge. In the first scenario, \citet[][their figure 8]{moles98} showed that this tail starts as a second arm South of NGC\,7318b, which first curls further South and West towards the elliptical galaxy NGC\,7318a (forming the `smiley face' in the {\it JWST} imaging of Fig.\,\ref{fig:acacarma}), and then continues in a more diffuse tail that stretched North. At the outer end of the tail, the molecular gas approaches the velocities of the CO(2-1) found in the ridge. Alternatively, the molecular gas in regions N1 and N2 may have undergone the same processes as the shocked material in the ridge. In this case, the shocked region could extend as a partial ring in the North-Western part of \SQ, possibly produced in the disk of NGC 7318b as it collides with the group \citep{xu05}.

\subsubsection{High-$v$ component (NGC\,7319 $\&$ SQ-A)}

The high-$v$ gas in NGC\,7319 and SQ-A is found near the systemic velocity of the \SQ\ group (Sect.\,\ref{sec:intro}). In the deep {\it JWST}/F090W imaging of Fig.\,\ref{fig:jwstaca}, faint diffuse emission appears to be stretching from region E2 in NGC\,7319 towards SQ-A (Fig.\,\ref{fig:jwstaca}). This could perhaps be a faint counter-tail to the prominent South-Eastern tail that stretches in the direction of NGC\,7320c (Sect.\,\ref{sec:intro}), with SQ-A located at the outer end of this putative counter-tail. However, it is also possible that SQ-A is a distinct region within \SQ\ that is not directly related to NGC\,7319. 

It is likely that SQ-A underwent interaction with the intra-group medium, given that it lies along the ridge of shocked molecular gas and has a high rate of star formation (Sect.\,\ref{sec:intro}). The peak of the CO(2-1) emission in SQ-A is spatially offset to the North from a second, fainter CO(2-1) component that coincides with the bright star formation evident in optical and near-infrared images (Fig.\,\ref{fig:SQA}). If the CO cloud was pre-existing at approximately its current location and velocity, then the burst of star formation is likely triggered where ram pressure or gravitational interaction with the intruder galaxy NGC\,7318b are taking place, meaning that the ram-pressure or gravitational effects would be propagating from South to North across SQ-A.

\subsection{Mass of molecular gas}
\label{sec:mass}

The mass of cold molecular gas in \SQ\ can be estimated from the CO(1-0) luminosity through $M_{\rm H_2}$\,$\sim$\,$\alpha_{\rm CO}$\,$\cdot$\,$L_{\rm CO(1-0)}^{\prime}$, with $\alpha_{\rm CO}$ the CO conversion factor \citep{bolatto13}.\footnote{This definition of $M_{\rm H_2}$ includes a 36 percent correction for helium.} Various studies have found $\alpha_{\rm CO}$ to range from $\sim$0.8 in ultraluminous infrared galaxies \citep{downes98} to $\sim$4 in both the Milky Way \citep{bolatto13} and high-$z$ starforming galaxies \citep{daddi10,genzel10}, although $\alpha_{\rm CO}$ values as low as $\sim$0.4 have been observed in outflowing gas \citep{pereira24}. Because of this uncertainty we give the molecular gas masses as a function of $\alpha_{\rm CO}$, where $\alpha_{\rm CO}$\,=\,1 corresponds to the quoted values in Table\,\ref{tab:results}. Although the CARMA data directly trace CO(1-0), their sensitivity is significantly lower than our ACA data of CO(2-1), hence we likely miss some of the fainter CO(1-0) emission. We therefore use the CO(2-1) data to derive the total molecular gas mass in \SQ. The integrated CO(2-1) flux density across \SQ\ is $\int_{v} S_{\rm CO} \delta {v}$\,$\sim$\,132\,$\pm$\,7 Jy\,\kms, as derived from the total intensity map in Fig.\,\ref{fig:mapregions}. To translate this into a molecular gas mass, we assume r$_{21}$\,=\,0.51\,$\pm$0.05, which is the average value weighted by the CO(1-0) luminosity (i.e., weight $w_i$\,=\,$L_{{\rm CO(1-0),}\,i}^{\prime}$/$\Sigma_{i}$$L_{{\rm CO(1-0),}\,i}^{\prime}$) for the regions listed in Table \ref{tab:results}. This results in a total molecular gas mass of $M_{\rm H_2}$\,=\,(1.4\,$\pm$\,0.1)\,$\times$\,10$^{9}$\,$\cdot$\,$\alpha_{\rm CO}$ M$_{\odot}$ for \SQ.

Molecular gas masses for the individual regions, estimated from CO(1-0), are given in Table \ref{tab:results}. When added together, these values recover (9.8\,$\pm$0.4)\,$\times$\,10$^{8}$\,$\cdot$\,$\alpha_{\rm CO}$ $M_{\odot}$, which is 70$\%$ of the total molecular gas mass that we derive from the integrated CO(2-1) emission, as quoted above. This suggests that $\sim$30$\%$ of the cold molecular gas that we detect with the ACA is distributed outside the individual regions that we studied in detail. Such CO emission is most prominent as the faint emission in the bridge (Fig.\,\ref{fig:bridge}), in the Southern inner arm of NGC\,7319 (East of region E3), at the tip of the Southern arm of NGC\,7318b (West and North-West of region E4), and in the Northern starforming region (SQ-A and SQ-A Extension in Fig.\,\ref{fig:SQA}). For example, the integrated CO(2-1) emission from the bridge in Fig.\,\ref{fig:bridge} has a flux density of $\int_{v} S_{\rm CO(2-1)} \delta {v}$\,=\,15.4\,$\pm$\,1.0 \jybmkms. This corresponds to $M_{\rm H_2}$\,=\,(2.3\,$\pm$\,0.5) $\times$\,10$^{8}$\,$\cdot$\,$\alpha_{\rm CO}$ M$_{\odot}$ (assuming r$_{21}$\,=\,0.36\,$\pm$0.06), which is 2.5$\times$ higher than the mass we estimate from the peak emission in region B1 (Table\,\ref{tab:fitting}). Similarly, the SQ-A Extension in Fig.\,\ref{fig:SQA} contains almost 30$\%$ of the CO(2-1) flux density of the peak emission in SQ-A (Table\,\ref{tab:fitting}).

\subsection{Excitation of molecular gas}
\label{sec:excitation}

Figure \ref{fig:propertyplot} includes a schematic overview of the CO luminosity ratio across \SQ, r$_{\rm 21}$\,=\,$L^{\prime}_{\rm CO(2-1)}$/$L^{\prime}_{\rm CO(1-0)}$. This ratio is a measure of the gas excitation, which depends on the gas temperature and density \citep[see, e.g.,][]{carilli13}. Thermally excited molecular gas typically approaches an r$_{\rm 21}$ of unity in the optically thick regime under local thermodynamic equilibrium (LTE) conditions, and can have r$_{\rm 21}$\,$>$\,1 if the gas is dense and optically thin \citep[e.g.,][]{eckart90}. Contrary, sub-thermally excited gas with lower density and temperature has r$_{\rm 21}$\,$<$\,1. A high gas excitation is often driven by a combination of high gas densities (typically at least 10$^{3.5-5}$ cm$^{-2}$; \citealt{liu21}) and energy input from photon-dominated regions (PDRs) associated with star formation, X-ray-dominated regions (XDRs) near AGN, and shocks \citep[e.g.,][]{riechers11,carniani19,pensabene21,esposito22}. A high gas excitation can also be the result of supersonic turbulence and high cosmic ray energy densities \citep{papadopoulos12}.

The matching beam-sizes of our ACA and CARMA data circumvent previous limitations that affected line-ratio analyses done for \SQ\ using the IRAM 30m single-dish telescope \citep{guillard12}. We derive r$_{\rm 21}$ values across \SQ\ that range from 0.29\,$-$\,0.86, with a weighted average of 0.51\,$\pm$0.05 (see Sect.\,\ref{sec:mass}). This average value is somewhat lower than the average line ratio of r$_{\rm 21}$\,=\,0.7 derived by \citet{leroy13} for nearby disk galaxies, which is based on observations by \citet{leroy09} that showed a spread of 0.48\,$<$\,r$_{\rm 21}$\,$<$\,1.06. It is also on the very low end of r$_{\rm 21}$ values found in other nearby galaxies \citep[see][their Table 2]{casoli91}. 

The highest gas excitation is found in the regions E1 and S1. S1 is the region closest to the main optical body of galaxy NGC\,7318b, while E1 corresponds to the innermost region of NGC\,7319. In E1, the molecular gas associated with the narrow component shows the highest excitation, r$_{\rm 21}$\,=\,1.36\,$\pm$\,0.64 (Table\,\ref{tab:fitting}). The broad high-velocity component in region E1, which likely represents either an outflow or turbulent gas along a dust lane (see Sect.\,\ref{sec:7319}), has a lower r$_{\rm 21}$\,=\,0.54\,$\pm$\,0.16. While this result may indicate lower gas densities or temperatures associated with the high-velocity component in NGC\,7319, it appears to contradict studies of infrared-bright galaxies and radio galaxies that found relatively high excitation for outflowing gas \citep{dasyra16,oosterloo17,lutz20,montoya24}.

Moderate gas excitation, in the range 0.4\,$\la$\,r$_{\rm 21}$\,$\la$0.7, is seen for most of the regions in \SQ. Given the uncertainties in r$_{\rm 21}$ (Table\,\ref{tab:results}), it is difficult to distinguish clear trends in r$_{\rm 21}$ among these regions. Nevertheless, in particular the ridge (R) shows gas with fairly uniform excitation, with an r$_{\rm 21}$\,$\sim$\,0.5 that is similar to the weighted average of all the regions in \SQ\ (Sect.\,\ref{sec:mass}). The molecular gas across NGC\,7319 (E) and NGC\,7318b (S) shows excitation conditions similar to those found in studies of other barred, interacting galaxies \citep{egusa22,maeda22}, while the value r$_{\rm 21}$\,=\,0.65\,$\pm$\,0.27 in the outermost region of the spiral arm in NGC\,7318b (region S4) is similar to r$_{\rm 21}$\,$\sim$\,0.7 found in a molecular cloud complex along the tidal arms of NGC\,3077 by \citet{heithausen00}. Therefore, the bulk of the molecular gas in \SQ\ shows excitation conditions as expected from interacting galaxies.

The lowest gas excitation is found in the bridge (B), SQ-A, and region N1. For the bridge, the low excitation, combined with the fact that the velocity dispersion is also significantly larger than in other regions (FWHM\,=\,184\,$\pm$\,16 \kms; Sect. \ref{sec:bridge}), suggests that the molecular gas has a lower density and is more turbulent than elsewhere in \SQ. SQ-A and N1 are the regions farthest away from the interacting galaxies. The low r$_{\rm 21}$\,=\,0.34\,$\pm$\,0.05 in SQ-A is particularly puzzling, given that SQ-A is a site of star formation, with likely an average radiation field as expected from PDRs. Previous single-dish observations by \citet{lisenfeld02} that used a matching beam-size for the CO(2-1) and CO(1-0) analysis found a higher line ratio of $r_{\rm 21}$\,=\,0.69\,$\pm$0.16 for SQ-A, but this was CO emission integrated across a region with an extent of roughly 20\,kpc, and thus included molecular gas on a much larger scale than our analysis. We warn that SQ-A is located around the FWHM of the primary beam of our CARMA data (Fig.\,\ref{fig:acacarma}), and the primary beam response of the heterogeneous CARMA array is complex (Sect.\,\ref{sec:carma}). Therefore, future CO(1-0) observations of this region should verify the low line ratios in SQ-A.

\subsection{High-velocity gas in NGC\,7319: relation to the radio source}
\label{sec:natureoutflow}

In Sect.\,\ref{sec:7319} we presented the gas kinematics in the center of NGC\,7319, and argued that the blueshifted, high-velocity component of CO (Fig.\,\ref{fig:AGN}) is likely either an outflow driven by the radio source, or turbulent gas at the location of a dust lane that coincides with the radio hot-spot. Based on {\it JWST} imaging, \citet{pereira22} showed that the radio jet is interacting with the surrounding interstellar medium (ISM). The velocity dispersion of the blueshifted CO(2-1) component, $\sigma$\,=\,123\,$\pm$\,8\,\kms\ (FWHM\,=\,289\,$\pm$\,18\,\kms), is in reasonable agreement with the value of  $\sigma$\,$\sim$150\,\kms\ found for the warm H$_{2}$ with {\it JWST} by \citet{pereira22}, who conclude that this molecular material decelerates the radio jet and causes the observed asymmetry in the radio structure \citep{xanthopoulos04}. 

At first glance, the kinematics of the blueshifted CO component are in apparent contrast with the outflow reported by \citet{aoki96}, which shows blueshifted velocities SW of the nucleus. However, this ionized outflow is detected predominantly further away from the nucleus, towards where we also find faint, blueshifted ($<$\,6400 \kms) CO(2-1) emission in the direction along the bridge (Fig.\,\ref{fig:bridge}). Moreover, a prominent blue wing, similar to the broad CO(2-1) component in region E1, is also visible in the [O\,{\small \sc III}] spectrum at the location of the peak in the optical continuum and 2$-$3 arcsec towards the NE \citep[][their Fig.\,8]{aoki96}. 

The high-velocity component of the CO in region E1 has a total H$_{2}$ mass of $M_{\rm H_2}$\,=\,(9.5\,$\pm$\,2.2)\,$\times$\,10$^{7}$\,$\cdot$\,$\alpha_{\rm CO}$ M$_{\odot}$, based on values given in Table \ref{tab:fitting}. If this high-velocity component represents an outflow driven by the radio jet, then we can estimate an outflow rate of the molecular gas. The spatial distance between the central dust component and the peak of the blueshifted component is $\sim$0.6\,kpc (Sect.\,\ref{sec:7319}), while the assumed average outflow velocity is $v_{\rm outfl}$\,=\,$\Delta v$\,+\,$\frac{1}{2}$FWHM\,$\sim$\,220\,\kms, with $\Delta v$ the difference in velocity between the peaks of the narrow and broad component. This results in an average mass outflow rate of $\dot{M}_{\rm outfl}$\,=\,(36\,$\pm$\,8)\,$\cdot$\,$\alpha_{\rm CO}$ M$_{\odot}$\,yr$^{-1}$. This is high enough to deplete the  molecular gas reservoir in region E1 in a few million years. 

\begin{figure*}
\centering 
\includegraphics[width=0.64\textwidth]
{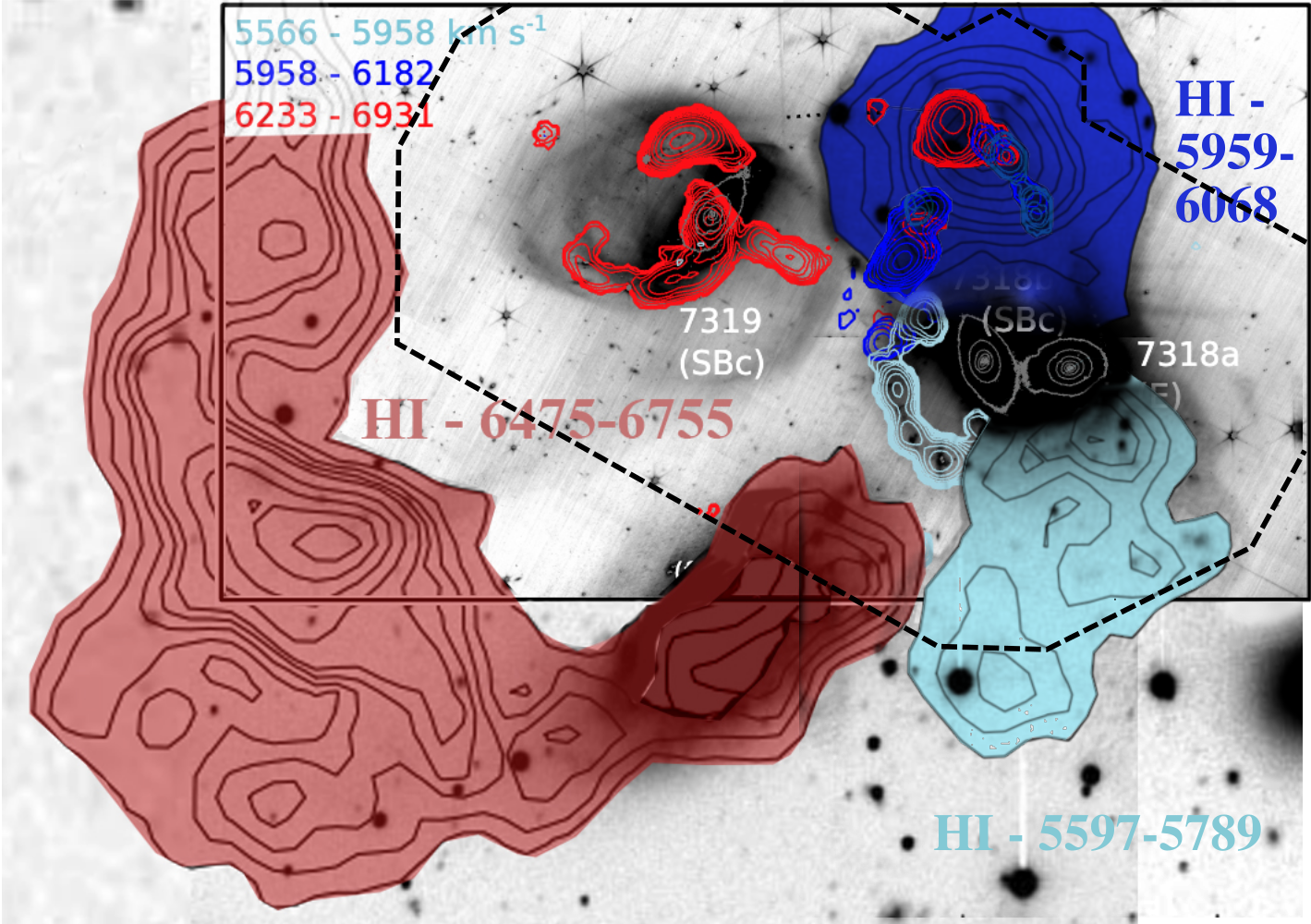}
\caption{Artistic rendering of Fig.\,\ref{fig:jwstaca} with the inclusion of VLA data of neutral hydrogen (\HI) gas from \citet{williams02}. The synthesized beam of these \HI\  observations is 19.4$^{\prime\prime}$\,$\times$\,18.6$^{\prime\prime}$, much larger than that of the ACA data. The color-coding of the \HI\ emission matches that of the CO(2-1), thus highlighting structures of neutral and molecular gas that are at approximately the same velocities, although the velocity dispersion of the CO(2-1) is somewhat higher than that of the \HI\ (see legend). The dashed polygon marks the field-of-view of our ACA data. For details of the \HI\ data we refer to \citet{williams02}.}
\label{fig:jwstaca+HI}
\end{figure*}

On the other hand, if the broad component represents turbulent gas associated with the region of the northern radio hot-spot and dust-lane, then its turbulent kinetic energy is $E_{\rm kin}^{\rm turb}$\,=\,$\frac{3}{2}$M$\sigma^{2}$\,$\sim$\,4.3\,$\times$\,10$^{55}$\,$\cdot$\,$\alpha_{\rm CO}$ erg, with $\sigma$\,=\,FWHM/2.35 the velocity dispersion. If we follow \citet{pereira22}, this is an order of magnitude larger than the bulk kinetic energy of the ionized and warm H$_{2}$ gas, and $\sim$6$\%$ (for $\alpha_{\rm CO}$\,=\,1) of the total energy of the radio jet. This adds to the scenario proposed by \citet{pereira22} that the radio source, which Northern lobe is brighter and shorter than the Southern lobe \citep{xanthopoulos04}, interacts with the dense gas and decelerates at the location where we find the broad CO component. Such an interaction between the jet and dense gas likely induces shocks that enhance the mid-infrared H$_{2}$ lines seen by \citet{pereira22} \citep[see][]{ogle10,guillard12b}. It likely also results in the brightening of the radio synchrotron emission, as previously found in other studies, such as those based on jet-CO alignments seen in high-$z$ radio galaxies \citep{lebowitz23,emonts23}, or on the large fraction of compact steep spectrum cores seen among low-$z$ starbursting radio galaxies \citep{tadhunter11}.

In either of the above scenarios (outflow vs. jet-ISM interaction), the broad CO component likely has a significant effect on both the gas content and the radio source in the central region of NGC\,7319. CO observations with higher spatial resolution are needed to further study the high-velocity component of CO in region E1.

\subsection{Scales of molecular gas: past, current, and future millimeter studies of Stephan's Quintet}
\label{sec:scales}

Our ACA CO(2-1) observations provide a global overview of the cold molecular gas across the inner $\sim$70 kpc of \SQ. Figure \ref{fig:jwstaca+HI} shows the CO(2-1) data compared to a map of 21-cm neutral hydrogen (\HI) gas, which stretches across a total extent of $\sim$4$^{\prime}$ or $\sim$100\,kpc\ \citep{williams02}. Overall, there is a poor correlation between the \HI\ and the CO. For NGC\,7319, the CO(2-1) is associated with the galaxy, while the large amounts of \HI\ (in red in Fig.\,\ref{fig:jwstaca+HI}) are detected far outside the galaxy \citep{shostak84, williams02}. In the North, no \HI\ is detected at the velocities of the CO in SQ-A. And a mis-match in location between the CO(2-1) and \HI\ is also seen South of NGC\,7318a/b. We note, however, that the field-of-view of our ACA observations is too small to image molecular gas across the full extent of the \HI\ emission (see Fig.\,\ref{fig:jwstaca+HI}). For example, CO emission was previously found in a region called SQ-B, which is at the outer end of the tidal tail that stretches from the southern part of NGC\,7319 in the direction of NGC\,7320c \citep{lisenfeld02,lisenfeld04}.

Our ACA data may also miss out widespread molecular gas on largest scales within our mosaic of \SQ. CO(1-0) observations performed by \citet{guillard12} with the 30m IRAM single-dish telescope revealed significantly broader line profiles than our ACA observations of CO(2-1) in various regions \citep[see also][]{lisenfeld02}. As mentioned in Sect.\,\ref{sec:kinematics}, this broad CO emission is found at velocities of broad H$\alpha$ emission due to shocks \citep{konstantopoulos14}. This could hint to the presence of turbulent cold gas spread on scales $\ga$14\,kpc ($\ga$30$^{\prime\prime}$), which is not picked up by our ACA observations (Sect.\,\ref{sec:acadata}). For example, across the ridge, single-dish spectra detected CO emission in the velocity range around 6500\,$-$\,7000 \kms\ \citep{guillard12}. This could perhaps be more diffuse (possibly post-shocked) gas. In our spectrum of region R2 (Fig.\,\ref{fig:spec1}), there is a slight hint of low-level emission between 6600 and 7000 \kms, but higher sensitivity is required to verify this faint emission and study its total extent. However, we warn that single-dish observations are affected by the primary beam response and side-lobe effects, and that bright emission from other regions in \SQ\ will leak into the single-dish spectra. For example, the single-dish spectrum of ridge-region R1 in \citet{guillard12} could contain some emission from the bridge. This makes a direct comparison between the IRAM single-dish results and our ACA results uncertain.

On smaller scales, our ACA data do not have the resolution or sensitivity to map cold molecular gas in individual cloud regions. \citet{appleton23} combined high-resolution observations of the warm and cold molecular gas in the brightest regions of the ridge, the bridge, and SQ-A. They found head-tail structures of warm H$_{2}$ stretching off clumps of cold CO(2-1), likely revealing the dissipation of kinetic energy due to shocks. These high-resolution data of the multi-phase molecular gas trace underlying physics that we cannot study with the ACA, but they miss a significant fraction of the overall molecular gas that our low-resolution data recover. A good example is SQ-A, where the high-resolution CO(2-1) data of \citet{appleton23} revealed that the secondary component ($v$\,$\sim$\,6650\,\kms) has a rich morphological structure in the region of bright emission seen with {\it JWST}. However, the main kinematic CO component in SQ-A (v\,$\sim$\,6864 \kms) is much brighter in our ACA data and its peak is off-set from the molecular structure seen by \citet{appleton23}. This suggest that our ACA data are revealing more widespread CO. Also in the regions of the bridge and ridge, \citet{appleton23} found a much richer morphological structure in CO, but only hints of low-surface-brightness CO emission along a filament, which is likely widespread CO that is detected at a much higher flux density in our ACA data of this region.

Overall, even though we are likely missing information about the molecular gas content on both the largest and the smallest scales, our ACA observations fill a critical gap in our understanding of the global distribution and kinematics of the cold gas. Complementary high-resolution observations of CO(2-1) with the ALMA 12m array are in progress to further study the shocking truth about the molecular gas across \SQ.

\section{Conclusions}

Our ACA data of CO(2-1) provided, for the first time with uniform sensitivity, a global overview of the morphology and kinematics of the cold gas across the inner $\sim$70 kpc of \SQ. Combined with CARMA data of CO(1-0), we studied the large-scale distribution, mass, and excitation conditions of cold molecular gas in the galaxies and intra-group medium. Our main conclusions are as follows:

\begin{itemize}

\item{CO(2-1) is found across more than a dozen regions, including along a ridge that crosses mid-way through the system, which previous studies found to contain shocked, multi-phase gas. In most regions, the global distribution of cold molecular gas matches the bright 10$\mu$m emission in {\it JWST/MIRI} imaging, which predominantly traces the H$_{2}$\,0-0\,S(3) line of warm molecular gas.}

\item{The global CO(2-1) kinematics reveal that there are three distinct kinematic structures in \SQ. A high-velocity structure near the systemic group velocity, which includes NGC\,7319 and the Northern star-forming region SQ-A; a mid-velocity structure that consists of the ridge; and a low-velocity structure that includes NGC\,7318b and a gas extension to the North, which could be either an extended arm of NGC\,7318b or part of the shocked gas in the system. The low- and mid-velocity structures overlap in velocity, linking the ridge to the intruder galaxy NGC\,7318b. There is a clear kinematic gap between the high- and mid/low-velocity structures.}

\item{A bridge of molecular gas covers the kinematic gap of $\sim$700\,\kms\  between the high- and mid/low-velocity structures, as low-surface-brightness emission stretches across $\sim$10\,kpc between NGC\,7319 and the ridge. The molecular gas associated with the brightest CO emission in the bridge has a high velocity dispersion (FWHM\,=\,184\,$\pm$\,16 \kms), indicating that the gas is turbulent.}

\item{Along the ridge, the CO velocity dispersion is lower than in other regions, with FWHM\,$\sim$\,25\,$-$\,65 \kms. This could be pre-existing cold gas, or gas that rapidly cooled and settled after being shocked by the intruding galaxy NGC 7318b. Our ACA data do not reveal broader CO emission at higher velocities, as previously seen in single-dish CO data and observations of H$_{2}$, H$\alpha$, and ionized gas.}

\item{We derive molecular gas masses for the various regions, and estimate a total molecular gas mass of $M_{\rm H_2}$\,=\,(1.4\,$\,$\,0.1)\,$\times$\,10$^{9}$\,$\cdot$\,$\alpha_{\rm CO}$ M$_{\odot}$ across \SQ.}

\item{The gas excitation varies across \SQ, ranging from r$_{\rm 21}$\,$\sim$\,0.3 in the bridge and SQ-A, to a rather uniform value of $\sim$0.5 along the ridge. The excitation of the gas associated with NGC\,7319 and NGC\,7318b is slightly larger, and consistent with values found in barred spiral galaxies (r$_{\rm 21}$\,$\sim$\,0.7). The gas excitation is highest in the central region of NGC\,7319 (r$_{\rm 21}$\,=\,1.36\,$\pm$0.64).}

\item{The radio-loud AGN of NGC\,7319 is associated with a broad, blueshifted CO component (FWHM\,=\,289\,$\pm$\,18 \kms). This broad component coincides with both the bright North-Eastern radio lobe and a central dust lane, while on larger scales it aligns with the gas in the bridge towards the South-West. This broad molecular component likely represents either a jet-driven outflow, or turbulent gas due to an interaction with the radio jet.}

\end{itemize}

This work with the ACA is part of a larger program with ALMA and {\it JWST} to study the various phases of molecular gas across \SQ\ (PI: Appleton; see also \citealt{appleton23}). Additional 12m observations of CO(2-1), combined with {\it JWST} integral-field spectroscopy of H$_{2}$, will be added in future work to zoom-in on molecular cloud complexes, with the goal of understanding the physics of the cold and warm molecular gas from the largest to the smallest scales across the intra-group medium of \SQ.

\begin{acknowledgments}
This paper makes use of the following ALMA data: ADS/JAO.ALMA$\#$2023.1.00177.S. ALMA is a partnership of ESO (representing its member states), NSF (USA) and NINS (Japan), together with NRC (Canada), MOST and ASIAA (Taiwan), and KASI (Republic of Korea), in cooperation with the Republic of Chile. The Joint ALMA Observatory is operated by ESO, AUI/NRAO and NAOJ. The National Radio Astronomy Observatory is a facility of the National Science Foundation operated under cooperative agreement by Associated Universities, Inc. 
This work is based on observations carried out with the CARMA telescope. Support for CARMA construction was derived from the Gordon and Betty Moore Foundation, the Eileen and Kenneth Norris Foundation, the Caltech Associates, the states of California, Illinois, and Maryland, and the NSF. Funding for CARMA development and operations were supported by NSF and the CARMA partner universities. This work is based on observations made with the NASA/ESA/CSA James Webb Space Telescope. The data were obtained from the Mikulski Archive for Space Telescopes at the Space Telescope Science Institute, which is operated by the Association of Universities for Research in Astronomy, Inc., under NASA contract NAS 5-03127 for JWST. These observations are associated with programs $\#$2732 and $\#$GO-3445. This research is based on observations made with the NASA/ESA Hubble Space Telescope obtained from the Space Telescope Science Institute, which is operated by the Association of Universities for Research in Astronomy, Inc., under NASA contract NAS 5–26555. These observations are associated with programs $\#$11502 and $\#$16123. UL acknowledges support by the research grants PID2020-114414GB-I00 from the Spanish Ministerio de Econom\'{i}a y Competitividad, from the European Regional Development Funds (FEDER) and the Junta de Andaluc\'{i}a (Spain) grants FQM108.
\end{acknowledgments}

\vspace{5mm}
\facilities{ALMA, CARMA, {\it JWST}, {\it HST}}

\software{CASA \citep{casa22}, MIRIAD \citep{sault95}}

\appendix

\section{Spectral analysis}
\label{sec:app_spectral}

\begin{figure}
\figurenum{A1}
\centering \includegraphics[width=0.75\columnwidth]{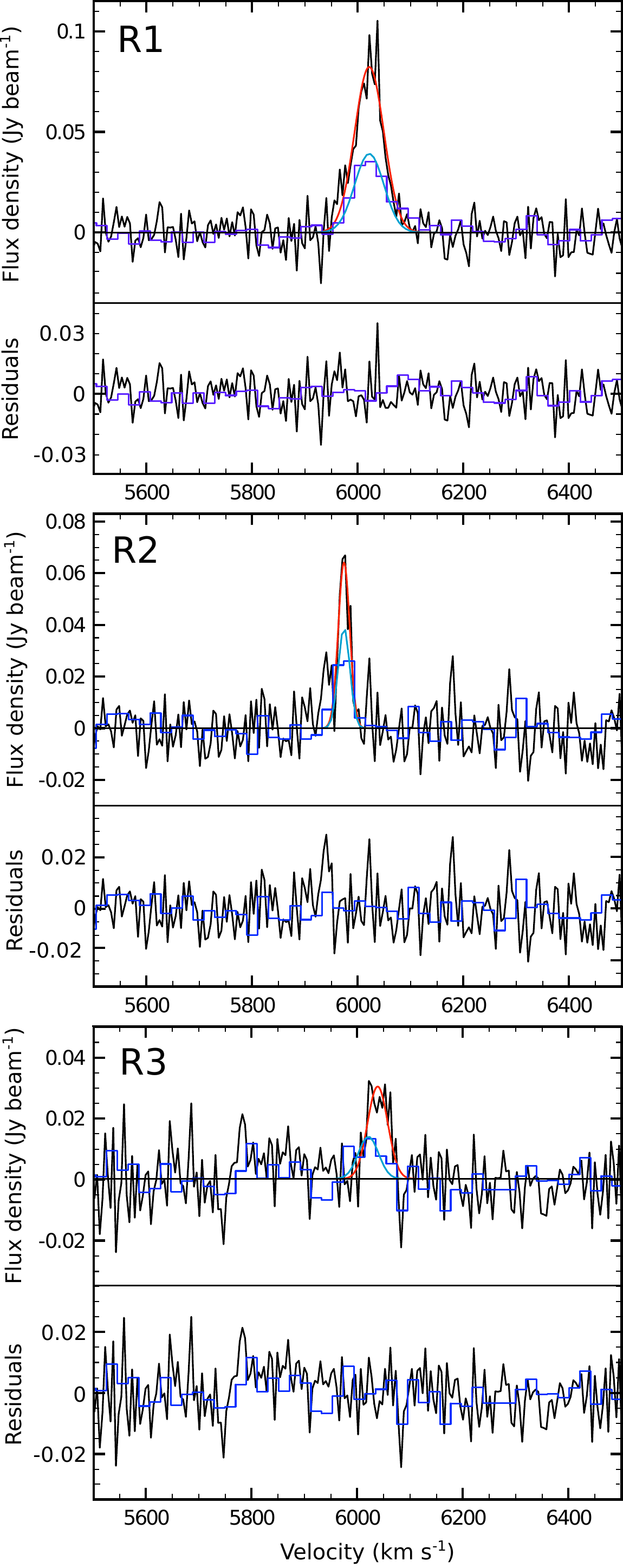}
\caption{Spectra of regions R1$-$R4. The black spectra are CO(2-1) observed with the ACA, while the blue histograms are CO(1-0) observed with CARMA. The red and light-blue lines are Gaussian fits to the ACA and CARMA spectrum, respectively. The flux densities of the ACA and CARMA spectra have been corrected for the primary beam response of the telescopes. For each of the CARMA spectra, the FWHM was constrained to match that of the corresponding ACA spectrum. The bottom panels show the residuals after subtracting the Gaussian models. Results of the fitting are shown in Table\,\ref{tab:fitting}.}
\label{fig:specR}
\end{figure}

In this appendix, we model the spectral lines of CO(2-1) and CO(1-0) extracted against the coordinates of the peak emissions in the various regions, as indicated in Fig.\,\ref{fig:mapregions}. The modeling is done by fitting Gaussian functions to the line profiles extracted from both the ACA and CARMA image cubes. This is used to derive the observed parameters listed in Table\,\ref{tab:fitting}. Because of the low signal-to-noise of the CARMA data, we constrained the FWHM of the CO(1-0) lines to be the same as the CO(2-1), thereby assuming that the CO(1-0) and CO(2-1) trace the same components of the molecular gas. Similarly, for multiple components within a CO(1-0) line, we constrained their velocity separation to that seen in CO(2-1). Spectra are shown in Figs.\,\ref{fig:specR}\,$-$\,\ref{fig:specNorth}, and results are summarized in Table \ref{tab:fitting}.

\begin{deluxetable*}{lccccccc}
\tablenum{A1}
\label{tab:fitting}
\tablecaption{Fitting of the ACA and CARMA spectra in the various regions, based on Figs.\,\ref{fig:specR}\,$-$\,\ref{fig:specNorth}. This table is an extended version of Table \ref{tab:results}. It separates out the different components for multi-component fits (single Gaussian fits are the same as in Table \ref{tab:results}).}
\tablehead{
\colhead{Region} & \colhead{R.A.} & \colhead{Dec.} &  \colhead{$v_{\rm opt}$} & \colhead{$FWHM$} & \colhead{$\int_{v} S_{\rm CO(2-1)} \delta {v}$} & \colhead{$\int_{v} S_{\rm CO(1-0)} \delta {v}$} & r$_{21}$ \\
\colhead{} & \colhead{J2000} & \colhead{J2000} & \colhead{(\kms)} & \colhead{(\kms)} & \multicolumn{2}{c}{(\jybmkms)} & \colhead{}
}
\startdata
R1 & 22$^{\rm h}$35$^{\rm m}$59.91$^{\rm s}$ & 33$^{\circ}$58$^{\prime}22.13^{\prime\prime}$ & 6022\,$\pm$\,3 & 65\,$\pm$\,4 & 5.73\,$\pm$\,0.25 & 2.67\,$\pm$\,0.18 & 0.54\,$\pm$\,0.06 \\
R2 & 22$^{\rm h}$35$^{\rm m}$59.58$^{\rm s}$ & 33$^{\circ}$58$^{\prime}$32.13$^{\prime\prime}$ & 5974\,$\pm$\,3 & 26\,$\pm$\,4 & 1.76\,$\pm$\,0.18 & 1.07\,$\pm$\,0.13 & 0.41\,$\pm$\,0.09 \\
R3 & 22$^{\rm h}$36$^{\rm m}$00.31$^{\rm s}$ & 33$^{\circ}$58$^{\prime}$01.13$^{\prime\prime}$ & 6039\,$\pm$\,3 & 45\,$\pm$\,6 & 1.45\,$\pm$\,0.18 & 0.67\,$\pm$\,0.14 & 0.55\,$\pm$\,0.19 \\
B1 & 22$^{\rm h}$36$^{\rm m}$02.08$^{\rm s}$ & 33$^{\circ}$58$^{\prime}$22.63$^{\prime\prime}$ & 6575\,$\pm$\,3 & 184\,$\pm$\,16 & 6.00\,$\pm$\,0.40 & 4.12\,$\pm$\,0.43 & 0.36\,$\pm$\,0.06 \\
B (full)$^{\dagger}$ & $-$ & $-$ & 6507\,$\pm$\,15 & 561\,$\pm$\,31 & 15.4\,$\pm$\,1.0 & $-$ & $-$ \\
SQ-A \hspace{6.5mm} a. & 22$^{\rm h}$35$^{\rm m}$58.94$^{\rm s}$ & 33$^{\circ}$58$^{\prime}$56.13$^{\prime\prime}$ & 6864\,$\pm$\,3 & 76\,$\pm$\,5 & 6.52\,$\pm$\,0.32 & 4.75\,$\pm$\,0.41 & 0.34\,$\pm$\,0.05 \\
SQ-A Ext. b.$^{\ddagger}$ & 22$^{\rm h}$35$^{\rm m}$58.86$^{\rm s}$ & 33$^{\circ}$58$^{\prime}$51.13 $^{\prime\prime}$ & 6672\,$\pm$\,3 & 50\,$\pm$\,10 & 1.85\,$\pm$\,0.26 & 1.28\,$\pm$\,0.28 & 0.36\,$\pm$\,0.09 \\
N1$^{\S}$ & 22$^{\rm h}$35$^{\rm m}$57.93$^{\rm s}$ & 33$^{\circ}$58$^{\prime}$47.13$^{\prime\prime}$ & 5956\,$\pm$\,3 & 25\,$\pm$\,4 & 1.79\,$\pm$\,0.19 & 1.54\,$\pm$\,0.24 & 0.29\,$\pm$\,0.08 \\
N2 & 22$^{\rm h}$35$^{\rm m}$57.33$^{\rm s}$ & 33$^{\circ}$58$^{\prime}$33.63$^{\prime\prime}$ & 5935\,$\pm$\,6 & 100\,$\pm$\,15 & 3.05\,$\pm$\,0.36 & 1.49\,$\pm$\,0.36 & 0.51\,$\pm$\,0.18 \\
S1 & 22$^{\rm h}$35$^{\rm m}$59.46$^{\rm s}$ & 33$^{\circ}$58$^{\prime}$08.63$^{\prime\prime}$ & 5882\,$\pm$\,6 & 118\,$\pm$\,16 & 3.42\,$\pm$\,0.34 & 0.99\,$\pm$\,0.21 & 0.86\,$\pm$\,0.27 \\
S2 $\#$1 & 22$^{\rm h}$36$^{\rm m}$00.03$^{\rm s}$ & 33$^{\circ}$57$^{\prime}$51.13$^{\prime\prime}$ & 5739\,$\pm$\,3 & 40\,$\pm$\,6 & 3.20\,$\pm$\,0.62 & 2.05\,$\pm$\,0.20 & 0.39\,$\pm$\,0.11 \\
\hspace{3.5mm} $\#$2 & 22$^{\rm h}$36$^{\rm m}$00.03$^{\rm s}$ & 33$^{\circ}$57$^{\prime}$51.13$^{\prime\prime}$ & 5788\,$\pm$\,13 & 89\,$\pm$\,15 & 2.73\,$\pm$\,0.54 & 0.88\,$\pm$\,0.32 & 0.78\,$\pm$\,0.44 \\
S3 $\#$1 & 22$^{\rm h}$35$^{\rm m}$59.78$^{\rm s}$ & 33$^{\circ}$57$^{\prime}$41.13$^{\prime\prime}$ & 5693\,$\pm$\,2 & 52\,$\pm$\,5 & 2.70\,$\pm$\,0.21 & 2.03\,$\pm$\,0.46 & 0.33\,$\pm$\,0.10 \\
\hspace{3.5mm} $\#$2 & 22$^{\rm h}$35$^{\rm m}$59.78$^{\rm s}$ & 33$^{\circ}$57$^{\prime}$41.13$^{\prime\prime}$ & 5830\,$\pm$\,5 & 118\,$\pm$\,11 & 4.47\,$\pm$\,0.32 & 1.78\,$\pm$\,0.46 & 0.63\,$\pm$\,0.21 \\
S4 $\#$1 & 22$^{\rm h}$35$^{\rm m}$59.18$^{\rm s}$ & 33$^{\circ}$57$^{\prime}$34.13$^{\prime\prime}$ & 5663\,$\pm$\,7 & 53\,$\pm$\,11 & 3.46\,$\pm$\,1.54 & 1.76\,$\pm$\,0.41 & 0.49\,$\pm$\,0.33 \\
\hspace{3.5mm} $\#$2 & 22$^{\rm h}$35$^{\rm m}$59.18$^{\rm s}$ & 33$^{\circ}$57$^{\prime}$34.13$^{\prime\prime}$ & 5710\,$\pm$\,11 & 90\,$\pm$\,9 & 5.97\,$\pm$\,1.44 & 1.89\,$\pm$\,0.55 & 0.79\,$\pm$\,0.42 \\
E1 $\#$1 & 22$^{\rm h}$36$^{\rm m}$03.64$^{\rm s}$ & 33$^{\circ}$58$^{\prime}$31.63$^{\prime\prime}$ & 6694\,$\pm$\,3 & 77\,$\pm$\,5 & 6.82\,$\pm$\,0.50 & 1.25\,$\pm$\,0.49 & 1.36\,$\pm$\,0.64 \\
\hspace{3.5mm} $\#$2 & 22$^{\rm h}$36$^{\rm m}$03.64$^{\rm s}$ & 33$^{\circ}$58$^{\prime}$31.63$^{\prime\prime}$ & 6619\,$\pm$\,11 & 289\,$\pm$\,18 & 9.78\,$\pm$\,0.69 & 4.49\,$\pm$\,1.02 & 0.54\,$\pm$\,0.16 \\
E2 $\#$1 & 22$^{\rm h}$36$^{\rm m}$04.17$^{\rm s}$ & 33$^{\circ}$58$^{\prime}$51.13$^{\prime\prime}$ & 6794\,$\pm$\,3 & 74\,$\pm$\,4 & 14.6\,$\pm$\,0.6 & - & - \\
\hspace{3.5mm} $\#$2 & 22$^{\rm h}$36$^{\rm m}$04.17$^{\rm s}$ & 33$^{\circ}$58$^{\prime}$51.13$^{\prime\prime}$ & 6705\,$\pm$\,4 & 83\,$\pm$\,9 & 7.42\,$\pm$\,0.71 & - & - \\
E3 & 22$^{\rm h}$36$^{\rm m}$04.33$^{\rm s}$ & 33$^{\circ}$58$^{\prime}$18.13$^{\prime\prime}$ & 6646\,$\pm$\,3 & 58\,$\pm$\,5 & 3.71\,$\pm$\,0.22 & - & - \\
\enddata
\tablecomments{$^{\dagger}$ Based on the integrated spectrum shown in Figs. \ref{fig:bridge} and \ref{fig:specB}.\\
$^{\ddagger}$ This is the secondary kinematic component of SQ-A, which brightness peaks the region defined as SQ-A Extension in Fig.\,\ref{fig:SQA}. This component is also seen in region N1 (Fig.\,\ref{fig:spec1}), but the fitting results only reflect the values obtained for the SQ-A Extension. For simplicity, the faint component {\sl b} is not included in Table \ref{tab:results}.\\
$^{\S}$ Not including component {\sl b}, which is analyzed in the region SQ-A Extension.}
\end{deluxetable*}

\begin{figure}[t!]
\figurenum{A2}
\centering \includegraphics[width=0.75\columnwidth]{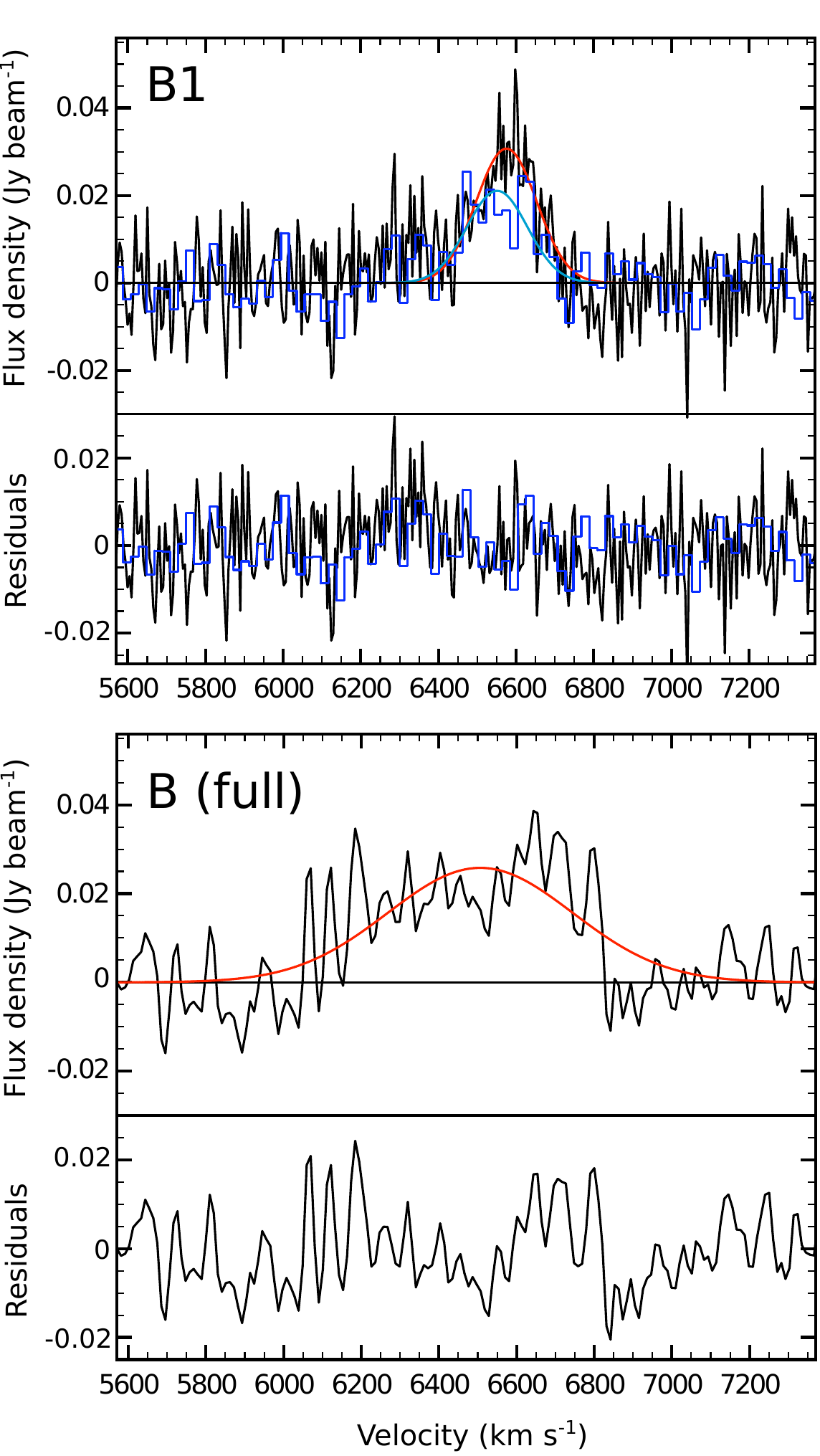}
\caption{Same as Fig.\,\ref{fig:specR}, but for region B1. The bottom panel shows the full region of the bridge, for which the data were binned by two channels and subsequently Hanning smoothed to visualize the broad signal (see also Fig.\,\ref{fig:bridge}). CARMA CO(1-0) results for the full bridge are unreliable, because the bridge stretches across a significant fraction of the primary beam, hence CO(1-0) results are only included for region B1.}
\label{fig:specB}
\end{figure}

\begin{figure}
\figurenum{A3}
\centering \includegraphics[width=0.75\columnwidth]{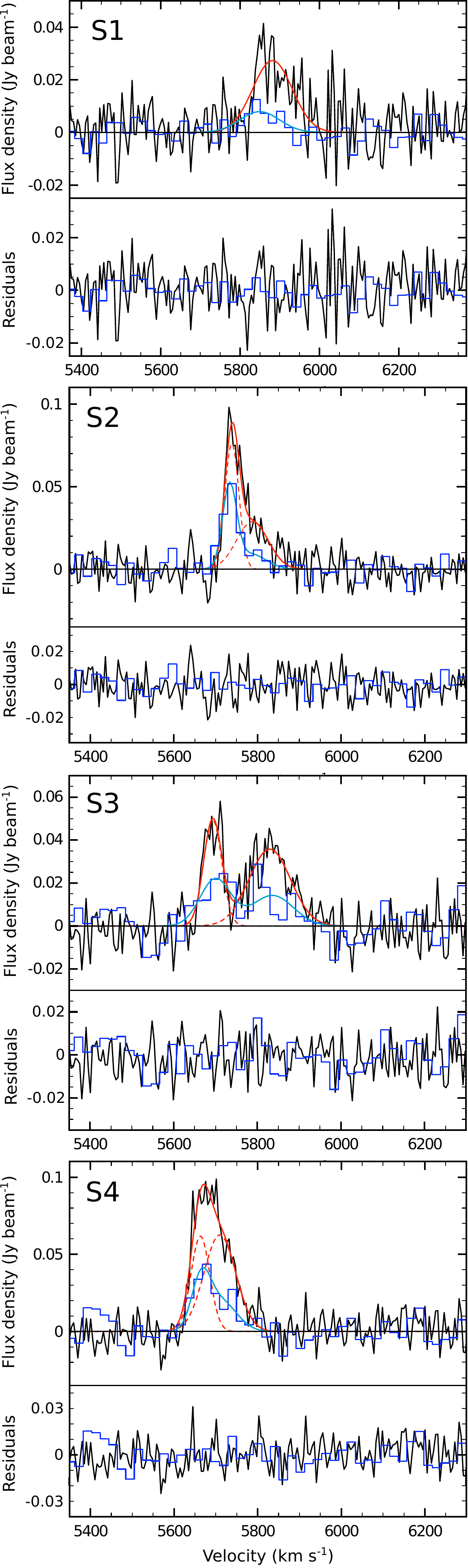}
\caption{Same as Fig.\,\ref{fig:specR}, but for region S1$-$S4. For regions S2$-$S4 two Gaussians (dashed red lines) are required to accurately model the ACA spectra. To fit the corresponding CARMA spectra, we also use two Gaussians, with their FWHM and velocity separation constrained to match those used in the fit of the ACA spectra. For clarity, the individual Gaussians are omitted from the CARMA spectra, and only the overall model fit is shown (light blue line).}
\label{fig:specS}
\end{figure}

\begin{figure}
\figurenum{A4}
\centering \includegraphics[width=0.75\columnwidth]{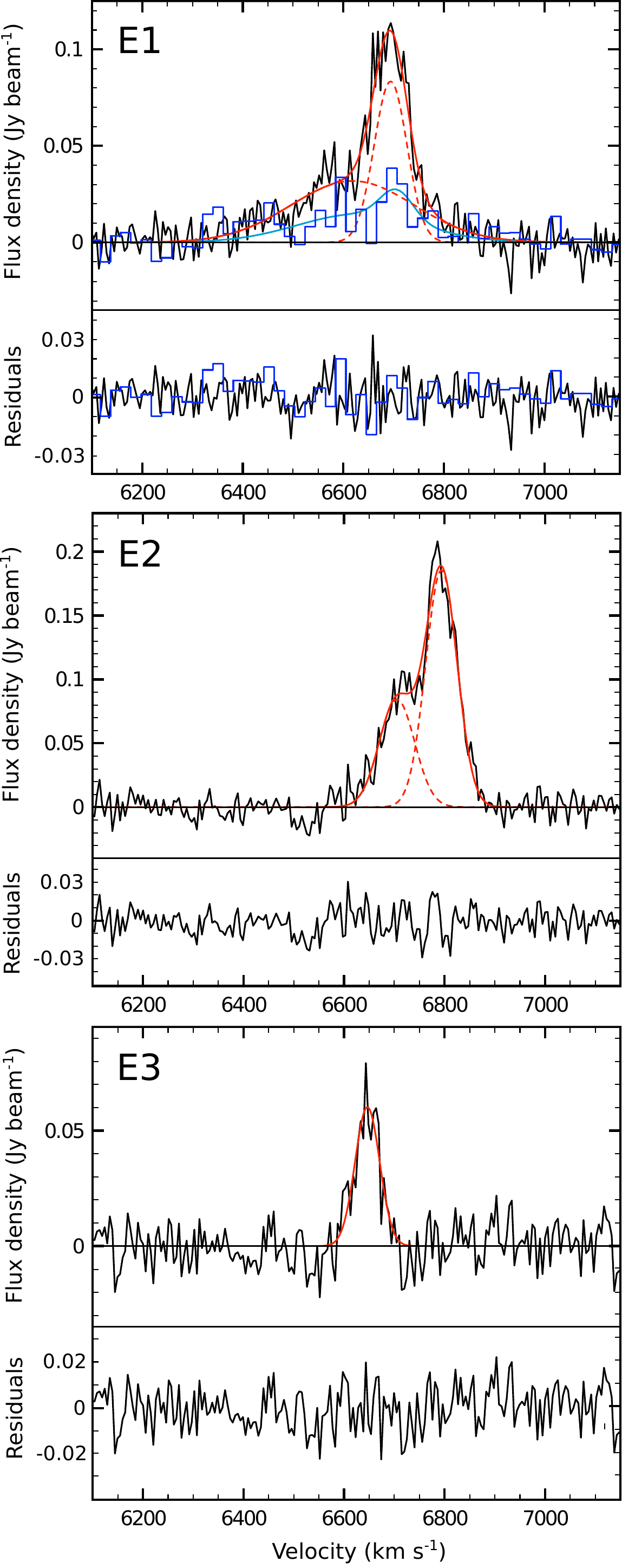}
\caption{Same as Fig.\,\ref{fig:specS}, but for regions E1$-$SE3. Regions E2 and E3 are located beyond the 35$\%$-level of the primary beam, hence the CARMA CO(1-0) results are unreliable and therefore not included in the analysis (Sect.\,\ref{sec:7319}).}
\label{fig:specE}
\end{figure}

\begin{figure*}
\figurenum{A5}
\centering 
\includegraphics[width=0.72\textwidth]{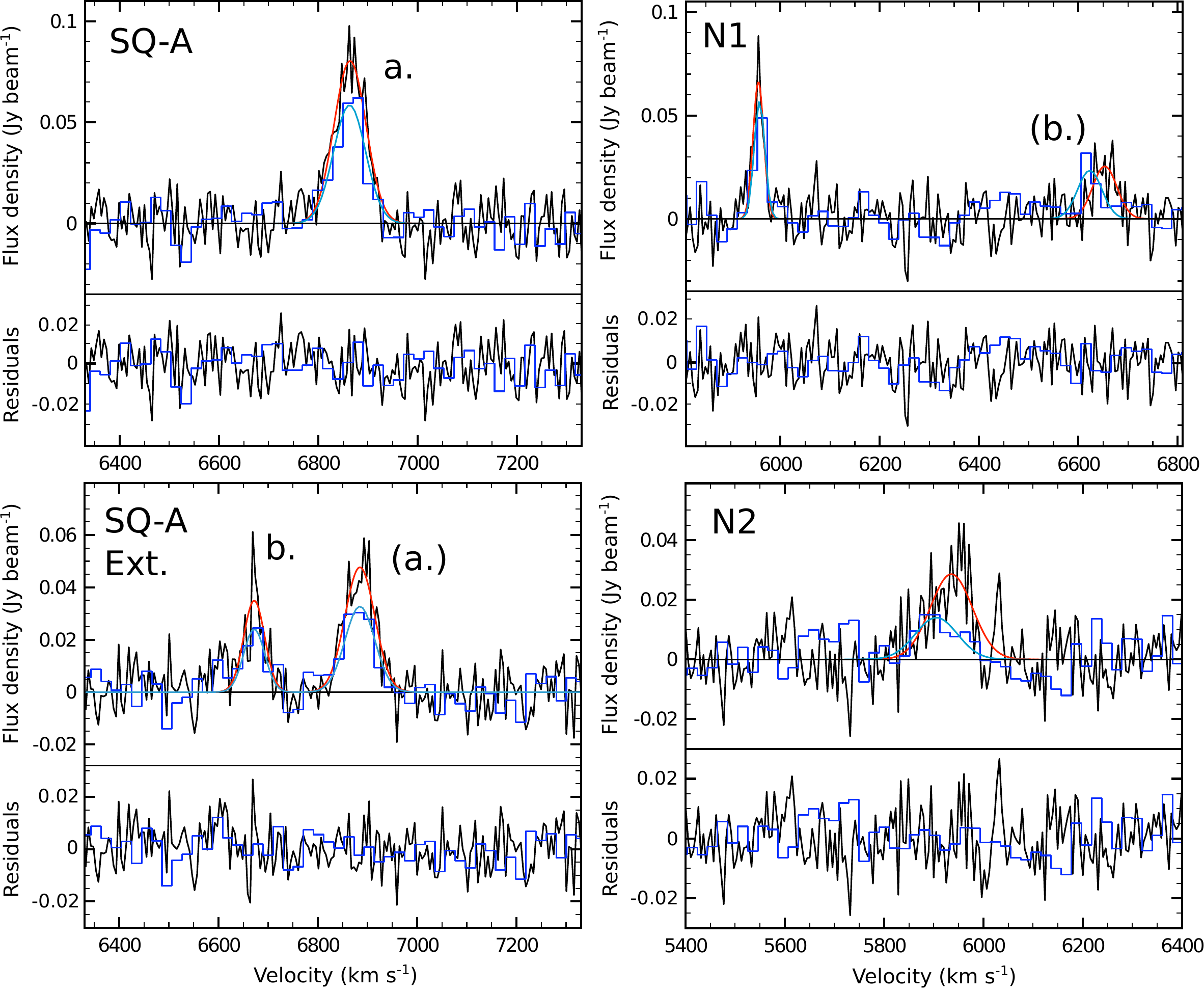}
\caption{Same as Fig.\,\ref{fig:specR}, but for regions in the North (SQ-A, N1, and N2). SQ-A Extension is the region just South of the CO peak in SQ-A. Components {\it a} and {\it b} are two kinematic components associated with the larger SQ-A region (Sect.\,\ref{sec:SQA}).}
\label{fig:specNorth}
\end{figure*}

%




\end{document}